\documentclass{aa}

\usepackage{graphicx}
\usepackage{subcaption}
\usepackage{txfonts}
\usepackage{bm}
\usepackage{hyperref}
\hypersetup{
    colorlinks,
    linkcolor=blue,
    citecolor=blue,
    urlcolor=blue
}

\newcommand{\Msolar}{\mathrm{M}_{\sun}}

\begin{document}

   \title{Joint cluster reconstructions}

   \subtitle{Combing free-form lensing and X-rays}

   \author{Korbinian Huber
          \inst{1,2}
          \and Céline Tchernin\inst{3}
          \and Julian Merten\inst{4}
          \and Stefan Hilbert \inst{1,2}
          \and Matthias Bartelmann\inst{3}
          }

   \institute{Ludwig-Maximilians-Universität, Fakultät für Physik, Universitäts-Sternwarte, Scheinerstraße 1, 81679 München, Germany
         \and
             Excellence Cluster Universe, Technische Universität München, Boltzmannstr. 2, D-85748, Garching, Germany
             \and 
             Center for Astronomy, Institute for Theoretical Astrophysics, Heidelberg University, Philosophenweg 12, 69120 Heidelberg,
Germany
\and INAF -- Osservatorio di Astrofisica e Scienza dello Spazio di Bologna, via Gobetti 93/3, 40129, Bologna, Italy\\
\email{korbinian.huber1@physik.uni-muenchen.de}
             }

   \date{Received ; accepted }

 
  \abstract
   {Galaxy clusters provide a multitude of observational data across wavelengths, and their structure and morphology are of considerable interest in cosmology as well as astrophysics.}
   {We develop a framework that allows the combination of lensing and non-lensing observations in a free-form and mesh-free approach to infer the projected mass distribution of individual galaxy clusters. This method can be used to test common assumptions on the morphology of clusters in parametric models.}
   {We make use of the lensing reconstruction code \textsc{SaWLens2}, and expand its capabilities by incorporating an estimate of the projected gravitational potential based on X-ray data that are deprojected using the local Richardson-Lucy method and used to infer the Newtonian potential of the cluster. We discuss how potentially arising numerical artefacts can be treated.}
   {We demonstrate the feasibility of our method on a simplified mock Navarro-Frenk-White (NFW) halo and on a cluster from a realistic hydrodynamical simulation. We  show how the combination of X-ray and weak lensing data can affect a free-form reconstruction, improving the accuracy in the central region in some cases by a factor of two.}
   {}

   \keywords{methods: numerical --
                galaxies: clusters --
                gravitational lensing: weak -- X-rays: galaxies: clusters
               }

   \maketitle
%

\section{Introduction}

Clusters of galaxies are extensively studied both en masse and in detail, and via a wide range of channels. In the context of large surveys, their total masses (depending on definition) are of considerable interest as they trace the exponential high-mass cut-off of the halo mass function and therefore are a sensitive probe of cosmological parameters \citep{allen:2011, kravtsov:2012}.
Individually, their matter distribution can shed light on numerous topics and in numerous ways, from the abundance of substructure, which can be used to test predictions from models of structure formation \citep{jauzac:2016} and the interplay of dark and baryonic matter in the formation of large-scale structure and galaxies \citep{dolag:2009,merten:2011} to cases where a galaxy cluster's strong lensing features can be used as a cosmic telescope, and thus offer data on otherwise unobservable distant galaxies \citep{kneib:2004} and,  via time delays, again on cosmology \citep{suyu:2017}. All of these factors rely on a  precise understanding of the morphology and structure of the galaxy cluster, so the accurate and precise reconstruction of the matter distribution in clusters  became an active field of research early on, and a variety of free-form methods have been developed \citep{schneider:1995, bradac:2005,diego:2007,liesenborgs:2010,umetsu:2013}.

Gravitational lensing and the X-ray emission of the intracluster medium (ICM) are both standard tracers of the mass distribution in clusters. Substantial efforts are being made to combine these and other probes in a statistically sound way. To give several recent examples, 
\citet{bonamigo:2018} used X-ray observations to infer the ICM mass distribution of three massive clusters and to use it as a fixed mass contribution in a subsequent strong lensing analysis to separate collisional and collisionless components. \citet{morandi:2012} and \citet{sereno:2017} use  lensing, X-ray, and thermal Sunyaev-Zel'dovich (SZ) observations of Abell 1835 and MACS J1206.2-0847, respectively, to jointly reconstruct the three-dimensional matter distribution.
Furthermore, gravitational lensing, as a fairly direct tracer of total mass, is used in several studies to calibrate mass observable relations for X-ray, SZ, or richness observations \citep{collaboration:2016, mantz:2016a,penna-lima:2017, bocquet:2018, dietrich:2019a, mcclintock:2019}, while \citet{sifon:2013} use the kinematics of cluster members to calibrate SZ scaling relations.
\\
In this work we present an attempt to combine a free-form lensing reconstruction framework with X-ray surface brightness data in a consistent way. This is the latest instalment in a series of papers, developing the lensing reconstruction framework to its current free-form and mesh-free state \citep{bartelmann:1996, merten:2009, merten:2016} on the one hand, and describing how to infer the gravitational potential from non-lensing tracers in a local, physically motivated way on the other \citep{reblinsky:2000, konrad:2013, sarli:2014, majer:2016, tchernin:2018}.

This paper is organised as follows. In Section \ref{sec:xdep}, after establishing the necessary lensing vocabulary, we review our method for estimating the gravitational potential based on X-ray data using the Richardson-Lucy deprojection algorithm, and point out the origins and a possible solution to the numerical artefacts that may arise. In Section \ref{sec:meth} we describe our combined reconstruction framework before  applying this method to mock clusters of increasing complexity in Section \ref{sec:test}. We  summarise our results in Section \ref{sec:conc}.


\section{Estimating the lensing potential with X-ray data}
\label{sec:xdep}

\subsection{Key concepts in lensing}
Gravitational lensing describes the deflection of light rays due to the gravitational potentials of massive objects along their path. As the core quantities of our method come from lensing, we  briefly summarise the main concepts used in this work  \citep[see e.g.][for an extensive review on the topic]{bartelmann:2010} in this section. The lens equation
 \begin{equation}
 \label{eq:lens}
 \boldsymbol\beta = \boldsymbol\theta - \boldsymbol\alpha\left(\boldsymbol\theta\right) 
 \end{equation}
 gives the relation between the true (angular) coordinates of a source $\boldsymbol\beta$, its observed coordinates $\boldsymbol\theta$, and the deflection angle due to gravitational lensing $\boldsymbol\alpha$. In the case of a thin lens, i.e. a negligible extent along the line of sight of all objects involved in a given lensing scenario, all lensing observables can be derived from the lensing potential, which is the line-of-sight projection of the gravitational potential of the lens 
\begin{equation}
 \psi \left(\boldsymbol\theta\right) = \frac{2}{c^2} \frac{D_\mathrm{ls}}{D_\mathrm{l}D_\mathrm{s}} \int \mathrm{d}z \Phi \left(D_\mathrm{l}\boldsymbol\theta, z\right), \label{eq:lpot}
 \end{equation}
and where $c$ is the speed of light and $D_\mathrm{l}$, $D_\mathrm{s}$, and $D_\mathrm{ls}$ are the angular diameter distances to the lens, the source and in between (see also Section \ref{sec:lens}).
To linear order, the mapping of an extended background source to the lens plane is then given by the lensing Jacobian
 \begin{equation}
 \mathcal{A}\left(\boldsymbol\theta\right) = \frac{\partial\boldsymbol\beta}{\partial\boldsymbol\theta} = \left( \delta_{ij} - \frac{\partial^2\psi\left(\boldsymbol\theta\right)}{\partial\theta_i\partial\theta_j} \right) .
 \end{equation}
We can then express the lensing quantities as combinations of derivatives with respect to the angular coordinates in the lens plane
\begin{eqnarray}
\boldsymbol \alpha &=& \boldsymbol\nabla_\theta \psi, \nonumber\\ \nonumber
\gamma_1 &=& \frac{1}{2}\left( \frac{\partial^2\psi}{\partial\theta_1^2} - \frac{\partial^2\psi}{\partial\theta_2^2} \right), \\\nonumber
\gamma_2 &=& \frac{\partial^2\psi}{\partial\theta_1\partial\theta_2},\\
\kappa &=& \frac{1}{2} \Delta_\theta \psi,
\end{eqnarray}
where we have defined the two components of the complex shear $\gamma=\gamma_1+i\gamma_2$ and the lensing convergence $\kappa$. In the weak lensing regime, the mapping results in slight distortions, for example turning originally circular images into ellipses. Assuming that the intrinsic ellipticities of background galaxies that are close by (in projection, not in physical distance) are uncorrelated and have no preferred orientation, we can relate their local average ellipticity $\langle \epsilon\rangle$ to the reduced shear
\begin{equation}
g = \frac{\gamma}{1-\kappa}
\end{equation}
via 
\begin{equation}
\langle \epsilon\rangle = \begin{cases}
g &\text{for $\left|g\right| \leq 1$}\\
1/g^\star &\text{for $\left|g\right| >1$}.
\end{cases}\label{eq:redshear}
\end{equation}

We note that once the projected cluster potential is known, other perhaps more tangible quantities can be derived from it. For instance, \citet{merten:2015} show how mass and concentration can be extracted from a free-form reconstruction of the lensing potential by explicitly and a posteriori assuming a parametrisation of the density. These derived mass values can then, for example, be related to cosmological parameters using established methods. We mention, however, that \citet{angrick:2009,angrick:2012} and \citet{angrick:2015}  have shown how cosmological information can be derived from cluster observations without reference to mass but rather to cluster potentials.

\subsection{Richardson-Lucy with correct amplitudes}
We briefly recap the main ideas of the Richardson-Lucy (RL) deprojection method. For more extensive accounts see \citet{lucy:1974,lucy:1994}, \citet{konrad:2013} and\citet{majer:2016}, among others.
For brevity and simplicity, here we  assume spherical symmetry and thus radial profiles, but the method can be extended to other symmetries as well \citep{reblinsky:2000,majer:2016}.

Assuming a three-dimensional field $f(r)$, a two-dimensional projection of this field can be expressed as
\begin{equation}
g(s) = \int \mathrm{d} r f(r) K(s|r),\label{eq:proj}
\end{equation}
where $r$ and $s$ denote a three- and two-dimensional radius, respectively, and $K(s|r)$ is the projection kernel, which for spherical symmetry reads
\begin{equation}
K(s|r) = \frac{r}{\sqrt{r^2-s^2}} \Theta (r^2-s^2),
\end{equation}
with the Heaviside step function $\Theta$.
Given the projected field $g(s)$, the RL deprojection algorithm infers locally (i.e. for each line of sight)  a reasonably free-form (i.e. no underlying model or functional form is assumed, only symmetry) estimate $\tilde f(r)$ of the three-dimensional field by iteratively maximising the objective functional 
\begin{equation}
Q[\tilde f] = H[\tilde f] + S[\tilde f],
\end{equation}
where 
\begin{equation}
H = \int \mathrm{d} s\, g(s) \ln{\tilde g(s)}
\end{equation}
is the log-likelihood for a reprojected estimate $\tilde g(s)$ and 
\begin{equation}
S = -\alpha \int \mathrm{d} r\, \tilde f(r) \ln{\frac{\tilde f(r)}{\chi(r)}} 
\end{equation}
an entropic regularisation with a smooth moving prior $\chi(r)$.
The algorithm preserves normalisation and also ensures, within a certain range of regularisation strengths $\alpha$,  non-negativity of all fields and kernels involved.

For reasons of internal consistency, fields and kernels need to be normalised with respect to their domains, but the correct amplitude and units for the resulting $\tilde{f}(r)$ can be restored as follows.
Assuming that the two-dimensional input field and the projection kernel are normalised via
\begin{eqnarray}
g_n(s) &=& \frac{1}{I_g} g(s),
\label{eq:gn}
\\
K_n(s|r) &=& \frac{1}{I_K(r)} K(s|r) 
\label{eq:Kn}
\end{eqnarray}
with
\begin{eqnarray}
I_g &=& \int \mathrm{d} s g(s),
\label{eq:Ig}
\\
I_K (r) &=& \int \mathrm{d} s K(s|r)
\label{eq:Ik},
\end{eqnarray}
then inserting Eqs.~\eqref{eq:Kn} and~\eqref{eq:proj} into Eq.~\eqref{eq:gn} and integrating over the two-dimensional domain yields
\begin{eqnarray}
\underbrace{\int \mathrm{d} s g_n(s)}_{=1} &=&\int \mathrm{d} r \frac{I_K(r)}{I_g}f(r) \underbrace{\int \mathrm{d} s K_n (s|r)}_{=1},
\end{eqnarray}
such that 
\begin{equation}
\int \mathrm{d} r \tilde f_n(r)=1,\qquad \tilde f_n(r) = \frac{I_K(r)}{I_g} f(r).
\end{equation}
 So we can recover the units and scale of the three-dimensional field via
\begin{equation}
\tilde f(r) = \frac{I_g}{I_K(r)}\tilde{f}_n (r)
\end{equation}
from the normalised output of the RL algorithm $\tilde{f}_n (r)$. Thus equipped with a deprojection method, we move on to connecting the deprojected quantity to the Newtonian potential.

\subsection{ICM physics}
\label{sec:rlxray}
In the case of X-ray observations, the observed two-dimensional data are counts of X-ray photons hitting a detector. Given average photon energy, exposure time, and effective area of the detector, this can be converted to the surface brightness $S_\mathrm{X}$, which in turn is the line-of-sight projection of the frequency integrated X-ray emissivity of the ICM
\begin{equation}
j_\mathrm{X} (r)= \mathrm{C} \rho^2(r) \sqrt{\mathrm{k_B}T(r)},
\end{equation}
where $\rho(r)$ and $T(r)$ are gas density and temperature, respectively, and the constant $\mathrm{C}$ is approximately $ 6.89\cdot10^{23} \,\frac{\mathrm{erg\,cm^3}}{\mathrm{g^2\,s}\sqrt{\mathrm{eV}}}$ for a fully ionised hydrogen plasma. To   estimate  the gravitational potential of the cluster from this, some assumptions about the state of the ICM have to be made. Here we opt for the following \citep[see][]{konrad:2013,tchernin:2018} :
\begin{itemize}
\item an ideal gas law;
\item a polytropic stratification;
\item an approximately hydrostatic equilibrium.
\end{itemize}
These assumptions are not necessarily expected to be valid for every cluster or across all radii;  nevertheless, they are often used. Studies suggest that the deviations from hydrostatic equilibrium are most relevant in the innermost cores and the outskirts of clusters, and  in recent mergers \citep{shi:2014, biffi:2016}. In addition,  polytropic stratification cannot necessarily safely be assumed for the centres of clusters as they may host cooling cores \citep{markevitch:1998}. The locality of the RL deprojection method, however, allows us to restrict ourselves to intermediate ranges in radius, which should minimise the effects of hydrostatic bias and localised cooling flows by neglecting the areas of the cluster where these biases may dominate.

With these assumptions the emissivity can be expressed as a function of the dimensionless gravitational potential $\varphi$ \citep{konrad:2013}
\begin{equation}
j_\mathrm{X}(r) = \mathrm{C}\rho^2_0 T^{1/2}_0 \varphi^{(3+\gamma)/(2\gamma -2)}(r),
\end{equation}
where $\rho_0$ and $T_0$ are reference values at a given radius $r_0$, $\gamma$ is the polytropic index, and the dimensionless potential is given by
\begin{equation}
\varphi (r) = -\frac{\gamma -1}{c_\mathrm{s,0}^2}\left(\Phi(r)-\Phi_\mathrm{cut}\right)
\end{equation}
with the sound speed at $r_0$
\begin{equation}
c_\mathrm{s,0}^2 = \gamma \frac{P_0}{\rho_0} = \gamma \frac{\mathrm{k_B}T_0}{\bar{m}}.
\end{equation}
The parameter $\Phi_\mathrm{cut}$ is introduced because the integration of the hydrostatic equation only fixes the Newtonian potential up to a constant.
Thus, the above-mentioned assumptions, an estimate of the polytropic index, and a measurement of  the sound speed, either via pressure and density or via the temperature of the ICM, are needed to infer the three-dimensional Newtonian potential $\Phi(r)$ from the observed X-ray surface brightness.


\subsection{Artefact treatment}
\label{sec:art}
We proceed by projecting the three-dimensional Newtonian potential from non-lensing observations to obtain an estimate of the lensing potential that can be incorporated in our reconstruction framework. This is  complicated, however,  by numerical artefacts arising from the  finite field of view of these observations.
\subsubsection{Origin of artefacts}
\begin{figure}
\begin{subfigure}[b]{\hsize}
\includegraphics[width=\hsize]{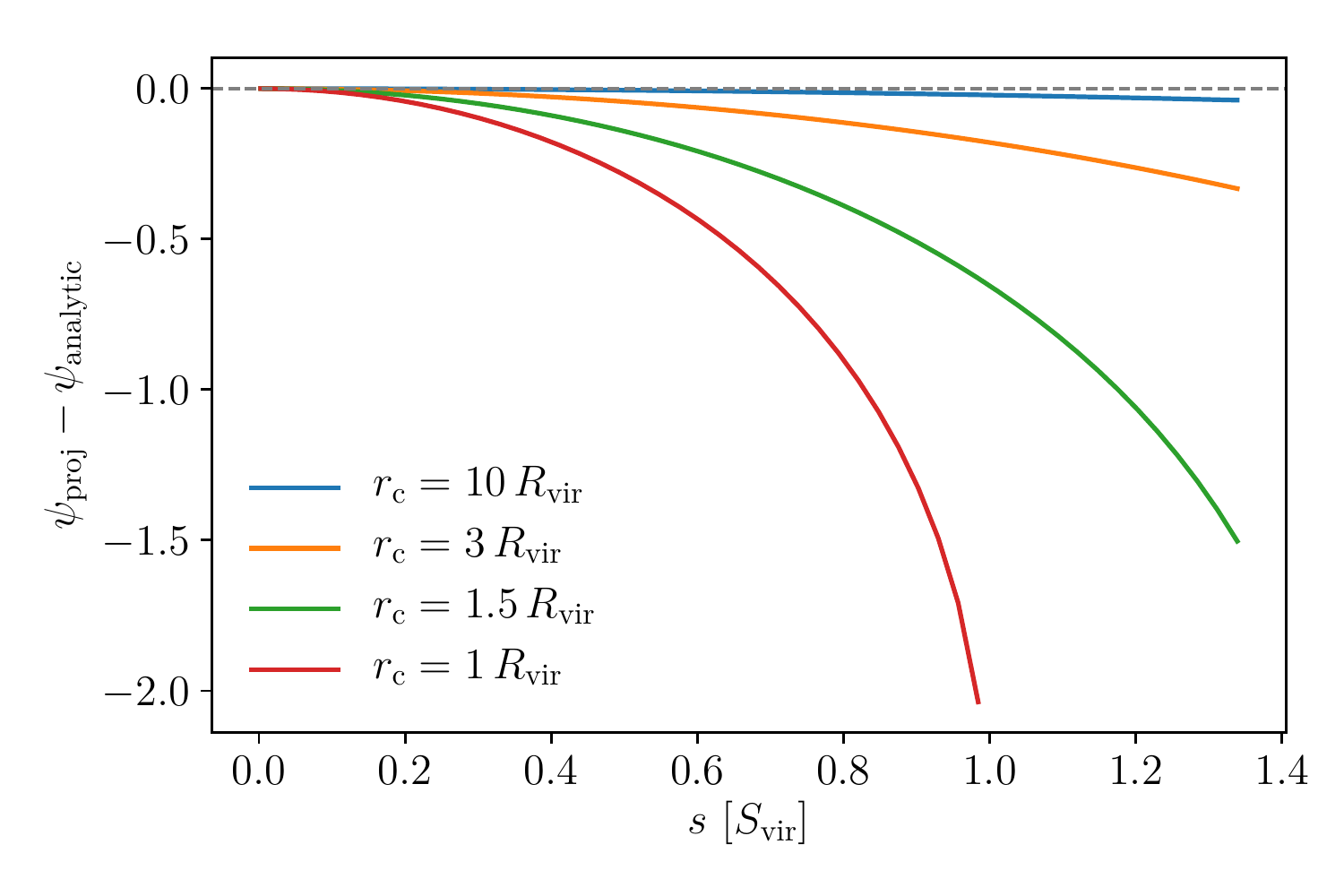}
\caption{potential}
\label{fig:projnfw}
\end{subfigure}
\hfill
\begin{subfigure}[b]{\hsize}
\includegraphics[width=\hsize]{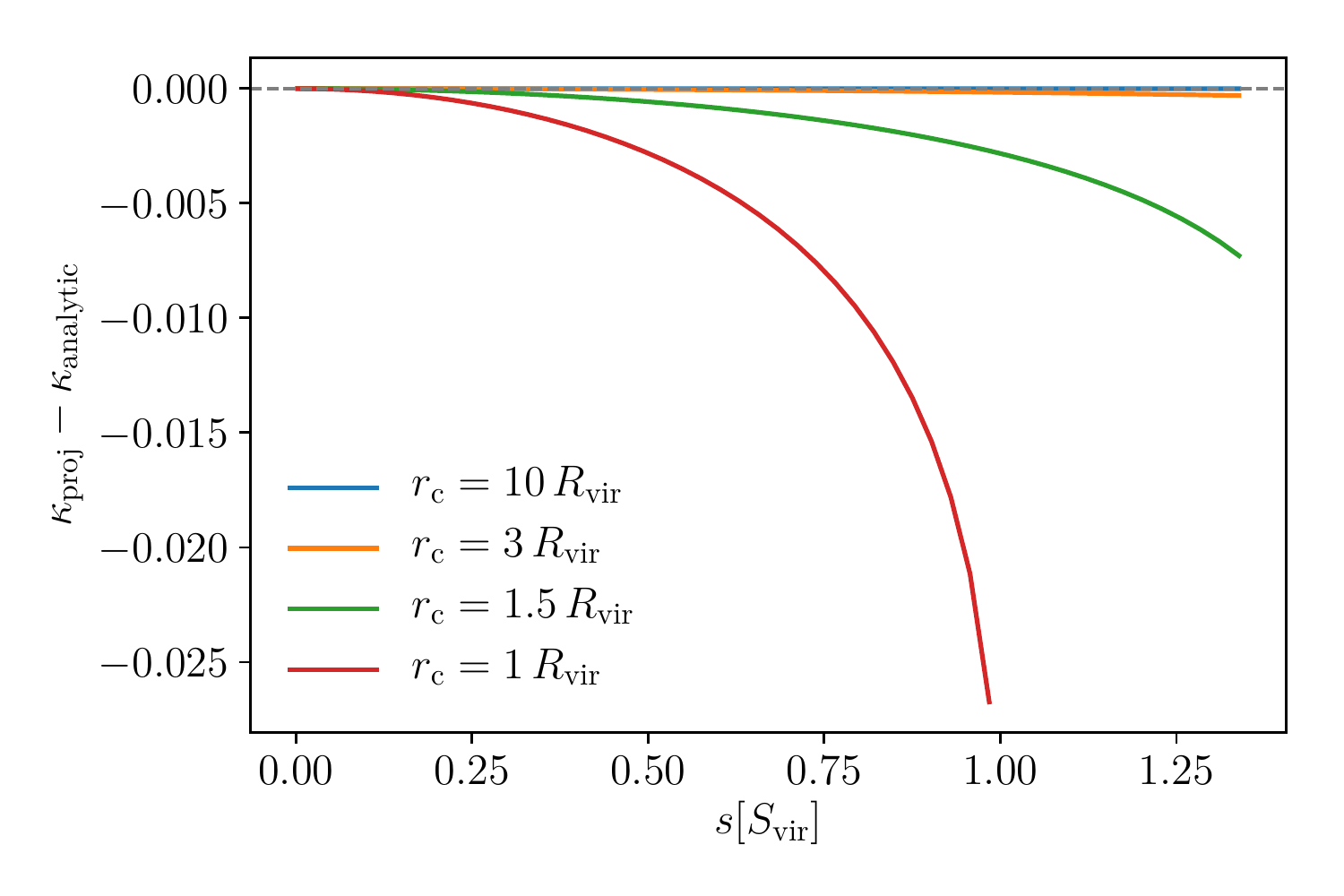}
\caption{density}
\label{fig:projnfw_dens}
\end{subfigure}
\caption{Projections of the gravitational potential (a) or the (rescaled) density (b) of a NFW halo, which are truncated at different multiples of the virial radius, vs. the analytic projected potential or density of that halo. All curves are shifted to coincide at the innermost radius.}
\end{figure}

The projection integral for the lensing potential under the assumption of spherical symmetry is formally  given by
\begin{equation}
\psi(s) = \frac{2}{c^2} \frac{D_\mathrm{ls}}{D_\mathrm{l}D_\mathrm{s}}\int_s^\infty \mathrm{d} r \frac{r}{\sqrt{r^2-s^2}} \Phi(r).
\end{equation}
However, the gravitational potential estimate from the emissivity, which is based on deprojected observations in a finite field of view, is only known up to a certain radius.  The projection integral at $r_\mathrm{data}$ can thus be truncated as
\begin{equation}
\psi \left(s\right) \approx \frac{2}{c^2} \frac{D_\mathrm{ls}}{D_\mathrm{l}D_\mathrm{s}} \int_s^{r_\mathrm{data}} \mathrm{d} r \frac{r}{\sqrt{r^2-s^2}} \Phi(r). \label{eq:lpot_p}
\end{equation}
The applicability of this approximation critically depends on $r_\mathrm{data}$. For large enough fields of view, the gravitational potential may drop to sufficiently small absolute values within $r_\mathrm{data}$ such that the unaccounted region does not significantly contribute to the integral and therefore can safely be neglected. However, if the field of view is  comparable to the virial radius of the cluster, these neglected contributions become increasingly important and simply cutting them away introduces artefacts that severely alter the radial curvature of the resulting lensing potential and systematically bias conclusions on the mass distribution of the lens. Figure~\ref{fig:projnfw} illustrates this effect. It shows the difference between the projected potential of a Navarro-Frenk-White (NFW) halo (Eq.~\eqref{eq:lpot_p}) and the analytical lensing potential. If the integral is truncated at the virial radius, the result significantly deviates from the truth for all radii. With increasing truncation radius, the differences become smaller and are restricted to increasingly large radii. This effect is less pronounced for functions that fall off more steeply, as can be seen in Fig.~\ref{fig:projnfw_dens} where the same procedure is shown for half the projected density compared to the analytically known convergence. Since the X-ray emissivity of the ICM is quadratic in the gas density and therefore falls off even more steeply, it is not surprising that truncation artefacts usually do not interfere with the RL deprojection algorithm even though it includes a reprojection in every iteration.

\subsubsection{Analytic continuation}
To mitigate these artefacts, we chose to introduce an analytic continuation of the 3D gravitational potential outside of $r_\mathrm{data}$. Several functional forms are possible. For this work we chose 
\begin{equation}
f_\mathrm{cont.}(r) = \frac{A}{r} + B,
\end{equation}
where $A$ and $B$ are fixed by  continuity conditions. The implied underlying assumption is   that there is no significant mass contribution outside of $r_\mathrm{data}$ and that therefore the potential in this radial range is dominated by the mass enclosed within $r_\mathrm{data}$. While this assumption most likely does not hold in a strict sense, tests with hydrodynamical simulations show that it is sufficient to reconcile reconstructed and true projected potential (see Section \ref{sec:hydrsim}). The modified projection integral then reads
\begin{align}
\psi \left(s\right) \approx \frac{2}{c^2} \frac{D_\mathrm{ls}}{D_\mathrm{l}D_\mathrm{s}} & \left[ \int_s^{r_\mathrm{data}} \mathrm{d} r \frac{r}{\sqrt{r^2-s^2}} \Phi(r) \right. \nonumber\\ & \left.+ \int_{r_\mathrm{data}}^{r_\mathrm{cut}} \mathrm{d} r \frac{r}{\sqrt{r^2-s^2}} f_\mathrm{cont}(r) \right],\label{eq:anacont}
\end{align}
where $r_\mathrm{cut}$ is an upper bound for the continuation, which needs to be introduced as most reasonable functional forms still formally diverge when integrated up to infinite radius.

\subsection{Sanity checks}
\label{sec:sanch}
As mentioned above, the parameters of the analytic continuation are set by  continuity conditions, i.e. by demanding that the values of $f_\mathrm{cont.}(r_\mathrm{data})$ and its first derivative coincide with the respective values in the deprojected profile. In order to ensure that especially the radial derivative for the latter is still driven by data rather than noise at $r_\mathrm{data}$ we employ a series of simple sanity checks.

\begin{figure}
\centering
\includegraphics[width=\hsize]{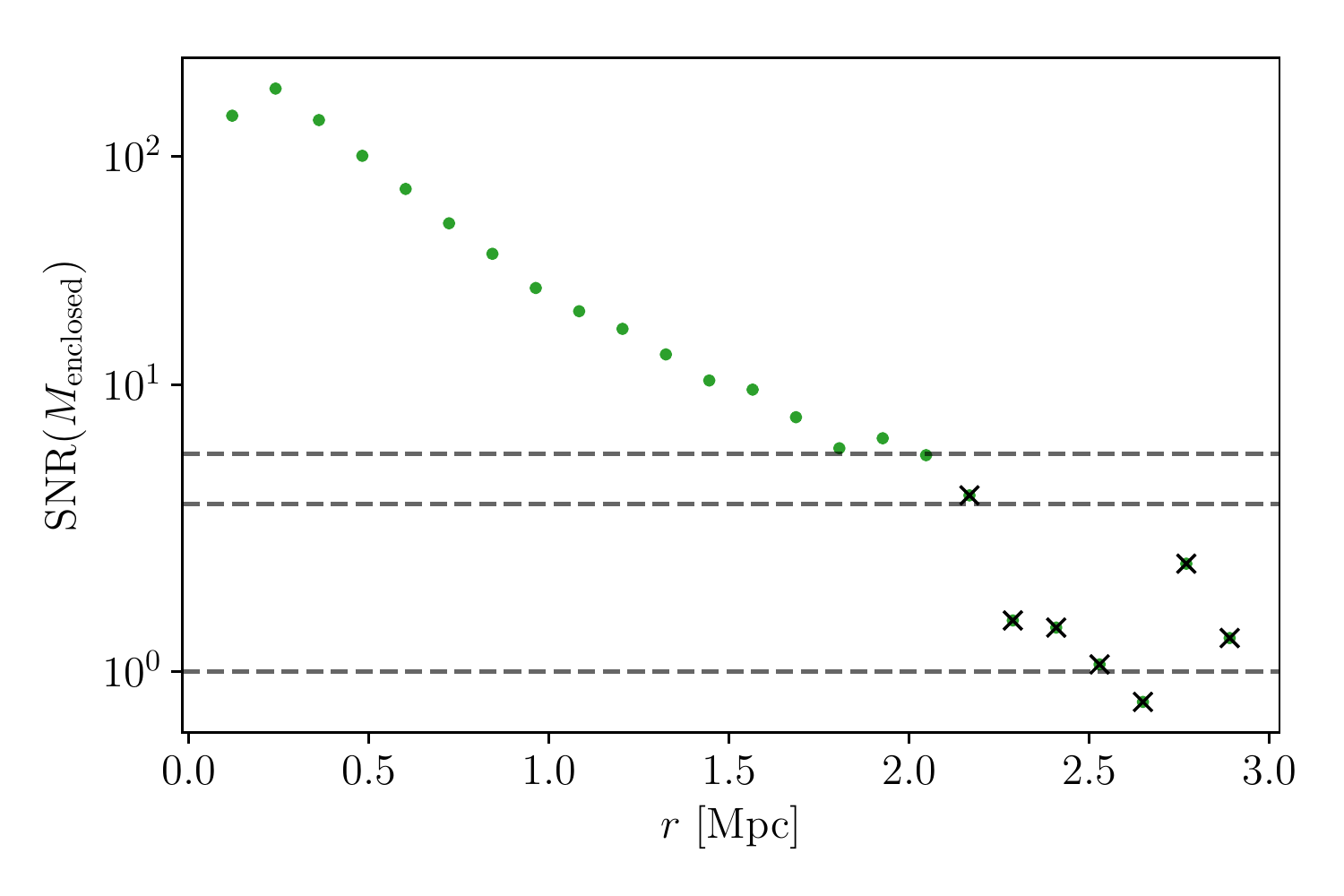}
\caption{Signal-to-noise ratio for the enclosed mass of a mock NFW halo as a function of radial bin. Horizontal lines denote values of  1, 3, and 5. Crossed-out bins fail the sanity checks and are excluded from the analysis. }
\label{fig:mmsnr}
\end{figure}

We estimate the signal-to-noise ratio for the enclosed mass as a function of radius
 \begin{equation}
 \mathrm{G} M_\mathrm{enclosed} (r) =  r^2\frac{\partial \Phi(r)}{\partial r}
 \end{equation}
from Poisson realisations of the X-ray data as this directly sets the slope of the continuation $A$. Figure~\ref{fig:mmsnr} shows an example for the mock NFW halo in Section \ref{sec:mock}.
Additionally we check for unphysical behaviour by testing the non-negativity of the Laplacian  $\Delta\Phi(r)$, which is proportional to the density via Poisson's equation.

If the outer bins fail to meet these criteria, the analytic continuation parameters would likely be set by random fluctuations and introduce new artefacts rather than correcting those induced by truncating the projection integral. Therefore, if necessary we cut the profile at the outermost radial bin where both conditions are still satisfied and start the continuation there. 


\section{Joint reconstruction framework}
\label{sec:meth}
\subsection{SaWLens2 in a nutshell} 
\textsc{SaWLens2} is a modular C++ code framework for free-form and mesh-free reconstructions of  projected gravitational potentials from observational data. Its core idea, to use a regularised maximum-likelihood estimator to reconstruct the lensing potential without relying on a specific mass model, was first presented in \citet{bartelmann:1996}. The maximum posterior reconstruction is found by directly solving the linear system of equations obtained from minimising the log-likelihood, which is done iteratively as necessitated by the non-linear dependence of some observables on the projected potential.
Previous implementations, based on adaptively refined but regular grid layouts as reconstruction domains, have been successfully applied to clusters in the CLASH sample \citep{postman:2012,merten:2015}. The latest mesh-free implementation based on strong and weak lensing constraints is described in detail in \citet{merten:2016}. We briefly summarise the main ideas here before focusing on ways to augment this framework with constraints from the ICM.

\subsection{Mesh-free support}
\textsc{SaWLens2} reconstructs the projected gravitational potential on a set of support nodes, which are not required to fulfil any sort of regularity or aspect ratio criteria. Thus, the reconstruction domain can easily be adapted to data sets that are distributed non-uniformly and on very different spacial scales, allowing for intuitive mask treatment and the combination of distinct data sets (e.g. a catalogue of weakly lensed background galaxies spanning several arc minutes with a set of strongly lensed multiple images only separated by tens of arc seconds). The obvious challenge is that interpolation and finite differencing schemes, which are required in free-form lens reconstructions, become less straightforward. Here, we make use of radial basis functions (RBF), a set of radially dependent functions centred on the nodes that make up the mesh-free domain. By expanding fields on the mesh-free domain in terms of these analytical RBFs, we can interpolate these fields to a different set of support nodes and define finite differencing stencils to obtain numerical derivatives.
For a detailed description of RBFs, mesh-free domains, and their application to lens reconstructions, see \citet{merten:2016}, \citet{fornberg:2011} and \citet{fornberg:2013} and references therein.
The handling of mesh-free domains, including interpolations and finite differences, is implemented in a publicly available C++ library\footnote{\url{https://bitbucket.org/jmerten82/libmfree}}.

\subsection{Lensing}
\label{sec:lens}

Equation~\eqref{eq:redshear}, restricted to the case of $\left|g\right| \leq 1$, sets the way shear measurements are incorporated in a \textsc{SaWLens2} reconstruction. After defining a set of reconstruction nodes, here chosen to be a subsample of sheared galaxy positions, ellipticities of galaxies in the vicinity of each node are averaged  to obtain an estimate of $\langle \epsilon\rangle$. This averaging process introduces covariances when neighbouring nodes have overlapping galaxy samples. This, together with  shape noise, is captured in the covariance matrix $C_\mathrm{WL}$ \citep[for details on its calculation see][]{merten:2016}.
The log-likelihood to be maximised by the reconstructed lensing potential then reads
\begin{equation}
\chi^2_\mathrm{WL}\left(\langle \epsilon\rangle|\psi\right) = \left( \langle \epsilon\rangle - \frac{1}{1-\kappa_\mathrm{p}} G \psi \right)^\mathrm{T} C_\mathrm{WL}^{-1} \left( \langle \epsilon\rangle -  \frac{1}{1-\kappa_\mathrm{p}} G \psi \right),
\end{equation}
where $G$ is the appropriate finite differencing operator for the respective component of the complex shear. We note that in order to maintain linearity in $\psi$, the value of $\kappa_\mathrm{p}$ is based on a prior and iteratively updated (see also Section \ref{sec:reg}).

\subsection{Including X-ray data}
Several ways to include a potential estimate  $\bar\psi_\mathrm{X}$ based on de- and reprojected X-ray data are conceivable, depending on the specific RL implementation and data available. In general, the respective log-likelihood term that enters the \textsc{SaWLens2} reconstruction will be of the form 
\begin{equation}
\chi^2_\mathrm{X} (\bar\psi_\mathrm{X}|\psi)= \left(\bar\psi_\mathrm{X} - P \psi \right)^\mathrm{T} C_\mathrm{X}^{-1} \left( \bar\psi_\mathrm{X} - P \psi  \right).\label{eq:RLchisq}
\end{equation}
Here $P$  denotes any operator that ensures compatibility of the coordinate basis of the model $\psi$ and that of the data-based $\bar\psi_\mathrm{X}$, and $C_\mathrm{X}$ denotes the covariance matrix of $\bar\psi_\mathrm{X}$.
If non-spherical symmetry assumptions (e.g. ellipsoidal) were used in the RL step, and thus $\bar\psi_\mathrm{X}$ were given for a grid with the resolution of the X-ray detectors CCD, $P$ would be designed to assign the reconstruction nodes of the mesh-free support to the corresponding pixels of $\bar\psi_\mathrm{X}$.

In the present case of spherical symmetry, $\bar\psi_\mathrm{X}$ is given as a radial profile obtained from azimuthally averaged X-ray surface brightness observations, and thus $P$ in this case is designed as a azimuthal averaging operator. It takes the form of an $m\times n$-matrix, where $m$ is the number of radial bins for $\bar\psi_\mathrm{X}$, and $n$ is the number of reconstruction nodes used. It has elements
\begin{equation}
P_{ij} = \begin{cases}
        N_i^{-1} & \text{if node }j\text{ falls in bin }i \\
        0 & \text{else},
        \end{cases}
\end{equation}
where $N_i$ is the number of nodes that lie within the radial bin $i$.
This way, the log-likelihood in Eq.~\eqref{eq:RLchisq} ensures that the reconstructed potential values $\psi$ on (azimuthal) average follow the profile $\bar\psi_\mathrm{X}$ while still leaving enough freedom  for local deviations from spherical symmetry required by lensing constraints, for example due to significant substructure or an overall non-spherical cluster morphology.

In order to get an estimate of $C_\mathrm{X}$ we create Poisson realisations of the observed photon count maps and subject them to the de- and reprojection process to arrive at realisations of the X-ray-based lensing potential, from which we can compute the corresponding covariance. We further use these realisations of the estimated potential to test if a Gaussian likelihood is applicable at all, and use the mean profile (which by construction is identical to the one based on the original data) in Eq.~\eqref{eq:RLchisq}.

\subsection{Regularisation}
\label{sec:reg}
Regularisation is required in a free-form framework to avoid overfitting. We adopt a two-level iteration scheme as described in \citet{bradac:2005}, \citet{merten:2009} and \citet{merten:2016}. Starting with a low number of initial support nodes allows us to  average over a comparably large number of neighbouring galaxies and hence significantly reduces shape noise. Subsequently increasing the number of nodes and reducing the number of galaxies in the averaging samples    increases resolution and allows to resolve smaller structures. On each resolution level we iteratively solve for $\psi$ while regularising on a moving prior in convergence and shear
\begin{eqnarray}
\chi^2_{\mathrm{reg,}\kappa} (\psi) &=& \left(\kappa_\mathrm{p} - K \psi\right)^\mathrm{T}H_\kappa \left(\kappa_\mathrm{p} - K \psi\right) \\
\chi^2_{\mathrm{reg,}\gamma} (\psi) &=& \left(\gamma_\mathrm{p} - G \psi\right)^\mathrm{T}H_\gamma \left(\gamma_\mathrm{p} - G \psi\right),
\end{eqnarray}
where $K$ is the finite differencing operator for (half) the 2D Laplacian and the diagonal matrices $H_{\kappa/\gamma}$ set the regularisation strength. With increasing resolution the result of the coarser gets interpolated to the finer domain and poses as the initial regularisation prior.

\subsection{Error estimation and goodness of fit}
\label{sec:errgof}
We analyse the normalised residuals to assess the quality of our reconstruction.
To this end we transform the residual vector
\begin{equation}
\bm r = \bm d - \bm m(\bm\psi),
\end{equation}
for the combined data set $\bm d$ and the corresponding combined response $\bm m(\bm \psi)$ into the eigenbasis of the combined inverse data covariance, in which the latter is diagonal, i.e.
\begin{equation}
\mathcal{T}^\mathrm{T}\mathcal{C}^{-1}\mathcal{T} = \mathrm{diag}(\lambda_1, \ldots, \lambda_N).
\end{equation}
Here $N$ is the total number of data points, $\lambda_i$ are the eigenvalues of the combined inverse data covariance, and $\mathcal{T}$ is the corresponding basis change operator. The normalised residuals are then
\begin{equation}
r_{\mathrm{n},i} = \left(\mathcal{T}^\mathrm{T}\bm r\right)_i \cdot \sqrt{\lambda_i},
\end{equation}
 which should follow a Gaussian distribution with zero mean and standard deviation $\sigma=1$, if all errors have been taken into account and regularisation has been chosen properly.

We employ re-sampling to estimate the uncertainties for the final reconstruction. We create a number of bootstrapped realisations of the shear catalogue and also bootstrap the X-ray-based lensing potential profile, by resampling the Poisson realisations created for the covariance estimation and averaging over the resampled set instead of the original to create the mean projected profile, but keeping the original covariance matrix.
The set of reconstructions obtained from these bootstrapped inputs is interpolated to the mesh-free domain of the default reconstruction and used to estimate the standard deviation of reconstructed quantities node by node. \footnote{A small fraction of bootstrap realisations returns entirely unphysical results (e.g. convergence values orders of magnitudes above the weak lensing regime) due to numerical instabilities. We discard the obviously corrupted cases before performing the error estimation.}

\section{Testing the method}
\label{sec:test}

\subsection{Mock NFW}
\label{sec:mock}

First we consider a simple cluster model where all assumptions are true by construction: a NFW halo with a polytropic ICM in hydrostatic equilibrium. We set the virial mass of the halo to $M_\mathrm{vir} = 5 \times 10^{14} h^{-1}\, \Msolar$ at redshift $z_\mathrm{l}=0.3$, assuming the concentration-mass-relation from \citet{seljak:2000} and \citet{klypin:2001}. 
The lensing properties of NFW halos are analytically known \citep[e.g.][]{golse:2002}, so we could easily create a mock shear catalogue  of roughly $6900$ randomly distributed galaxies, where we added Gaussian shape noise with $\sigma=0.3$.
We followed the formalism of \citet{komatsu:2001} to model the ICM in hydrostatic equilibrium with a polytropic stratification with index $\gamma=1.13$. We created a surface brightness map with the same field of view as the shear catalogue and assumed a mean photon energy of $1.5\,\mathrm{keV}$, an exposure time of $3000\,\mathrm{s,}$ and a detector with an effective area of $500\,\mathrm{cm}^2$. All defining properties of the halo and the mock observations are summarised in Table~\ref{tab:nfw}.

\begin{table*}
\centering
\begin{tabular}{c|c|c}
\hline \hline
Halo & Lensing & X-ray\\ 
\hline 
$M_\mathrm{vir} = 5.0\cdot10^{14} \Msolar h^{-1}$ & FoV$=(21\,\mathrm{arcmin})^2$ & FoV$=(21\,\mathrm{arcmin})^2$\\
$R_\mathrm{vir} =  1.71\,\mathrm{Mpc}$ & $z_\mathrm{s}=1.0$ & $\hbar\bar{\omega} =1.5\,\mathrm{keV}$\\
$c= 4.71$& $n_\mathrm{gal}\approx 20\,\mathrm{arcmin}^{-2}$ & $t_\mathrm{exp}=3000\,\mathrm{s}$\\
$z_\mathrm{l}=0.3$ & $\sigma=0.3$ & $A_\mathrm{eff}=500\,\mathrm{cm}^2$\\
$f_\mathrm{gas}=0.2$ & ~ & ~ \\
$\gamma = 1.13$ & ~ & ~\\
\hline 
\end{tabular} 
\caption[NFW mock parameters]{Defining parameters for the NFW halo and its Mock observation.}
\label{tab:nfw}
\end{table*}

We added Poisson noise to the resulting X-ray surface brightness before creating $20\,000$ Poisson realisations of the surface brightness profile in $25$ linearly spaced radial bins. 
We deprojected these realisations with the RL algorithm to determine the Newtonian potential profile, using the correct polytropic index and temperature of $8.1\, \mathrm{keV}$. 

\begin{figure}
\centering
\includegraphics[width=\hsize]{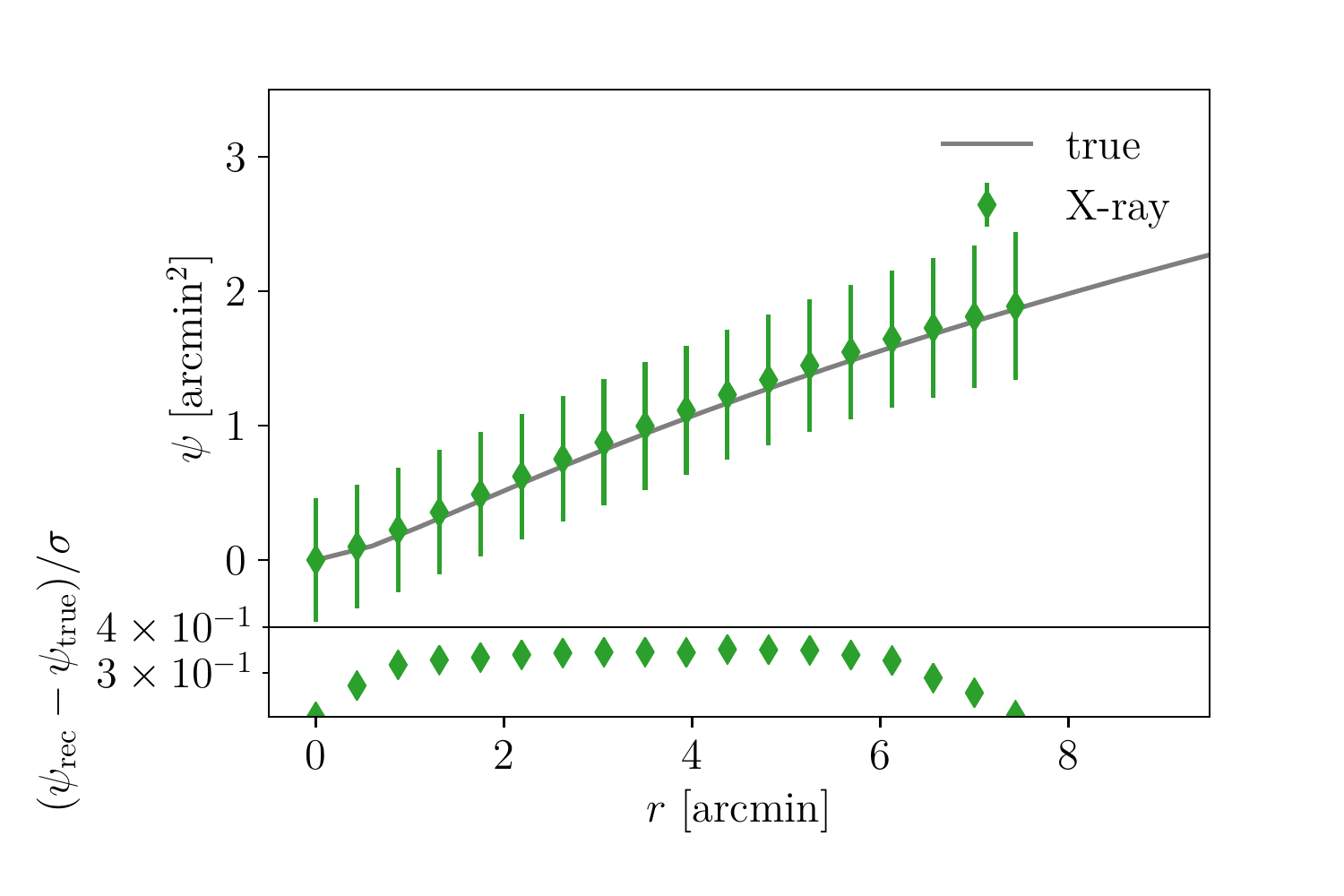}
\caption{Lensing potential of a NFW mock cluster, obtained from projecting the X-ray-based estimate of the Newtonian potential. }
\label{fig:nfw_projpsi}
\end{figure}

Based on the realisations of $\Phi_\mathrm{X}(r)$, we estimated the signal-to-noise ratio of the enclosed mass as a function of radius, as discussed in Section \ref{sec:sanch} and depicted in Fig.~\ref{fig:mmsnr}. We also computed the Laplacian of the potential as a proxy for the total density. We discarded all bins that fail our checks, i.e. where the Laplacian is negative or where the signal-to-noise ratio falls below $3$. The thus truncated potential profile was subsequently projected along the line of sight, using the analytic continuation according to Section \ref{sec:art} up to a cut-off radius of three times the truncated data radius.
The reconstructed lensing potential profile is compared to the true profile in Fig.~\ref{fig:nfw_projpsi}. The error bars depict the standard deviation within the sample of realisations, and their large size in comparison to the scatter in the data points in part reflects the high degree of correlation between bins. 

Even without knowing the true lensing potential, an appropriate cut-off radius for the analytical continuation can be found by gradually increasing it from the truncated data radius. If the cut-off radius is too small, the projection artefacts manifest themselves in deviations in the radial curvature of the potential at increasing radii (with increasing cut-off, see Fig.~\ref{fig:projnfw}). These curvature deviations lead to ring-like artefacts on the reconstructed convergence map. This of course does not pin down the proper cut-off radius entirely, but gives a lower limit. 

The realisations of $\psi_\mathrm{X}$ are Gaussian distributed and we scale all realisations to a reference redshift of $z_\mathrm{inf}=20\,000$ and use them to compute the inverse covariance matrix of the projected potential profile. 

With in place, we ran \textsc{SaWLens2}, once  using the shear data alone and once using both shear data and X-ray data. In both cases the mesh-free support was defined by subsampling the shear catalogues' galaxy positions. We chose three resolution levels, with $300$, $500$, and $800$ nodes, leading to a final resolution of roughly $0.5\,\mathrm{arcmin}^2$ per node. All parameters of the reconstruction are collected in Table~\ref{tab:nfw_rec}. The normalised residuals, as discussed in Section \ref{sec:errgof}, as well as further diagnostics of this and the following test case are presented in Appendix \ref{ap:test}.

\begin{table}
\centering
\begin{tabular}{l|ccc}
\hline \hline
Nodes (nominal) & $300$ & $500$ & $800$\\ 
Nearest neighbours & $33$ &$21$ & $12$\\
Regularisation strength &$200$&$400$&$600$\\
\hline 
\end{tabular} 
\caption[NFW reconstruction parameters]{List of parameters defining the outer loop iterations for the \textsc{SaWLens2} reconstruction of a NFW halo. The algorithm has some leeway when subsampling the shear catalogue to set up the mesh-free domain, so the actual numbers of nodes deviates slightly from the nominal value. The regularisation strength is set to be the same for  the regularisation against the convergence and for the regularisation against the shear.}
\label{tab:nfw_rec}
\end{table}

\begin{figure}
\centering
\includegraphics[width=\hsize]{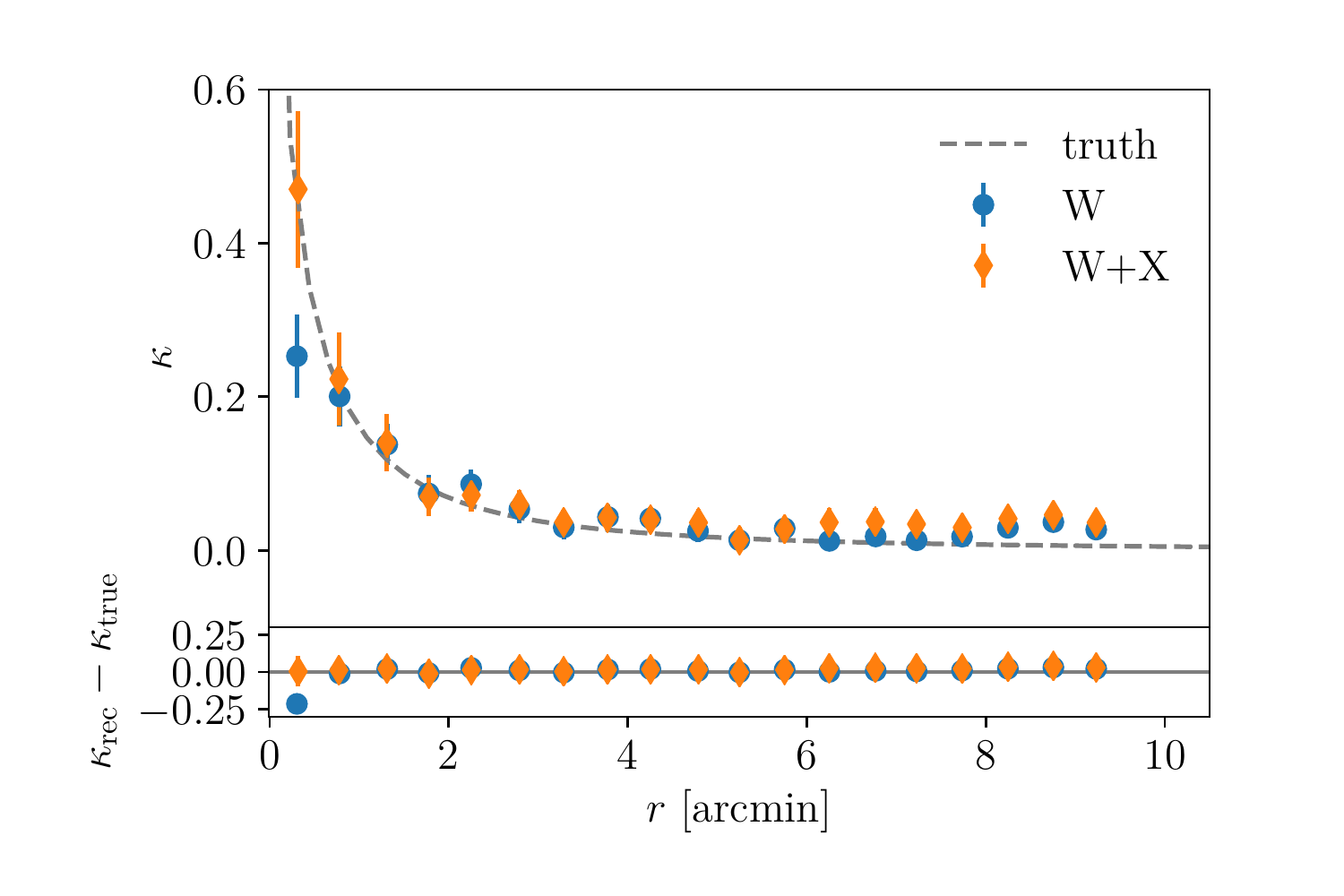}
\caption{Reconstructed convergence profiles for a NFW mock cluster (errors inferred from averaging over bootstrapped samples).}
\label{fig:nfwkappaprofdefBSerr}
\end{figure}

\begin{figure*}
\begin{subfigure}[b]{0.49\hsize}
\centering
\includegraphics[width=0.9\hsize]{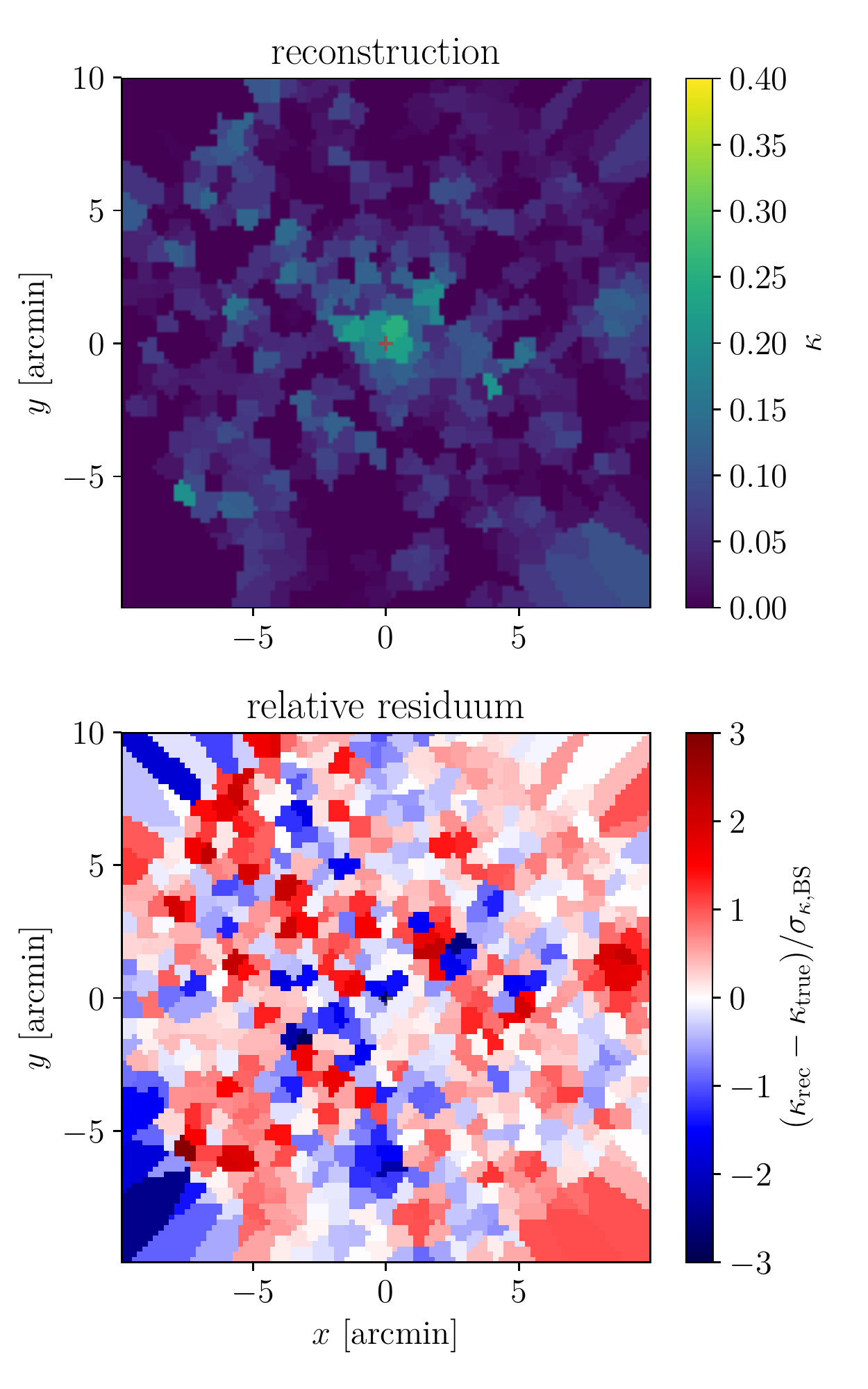}
\caption{weak lensing alone}
\label{fig:nfwmapwdefbs}
\end{subfigure}
~
\begin{subfigure}[b]{0.49\hsize}
\centering
\includegraphics[width=0.9\hsize]{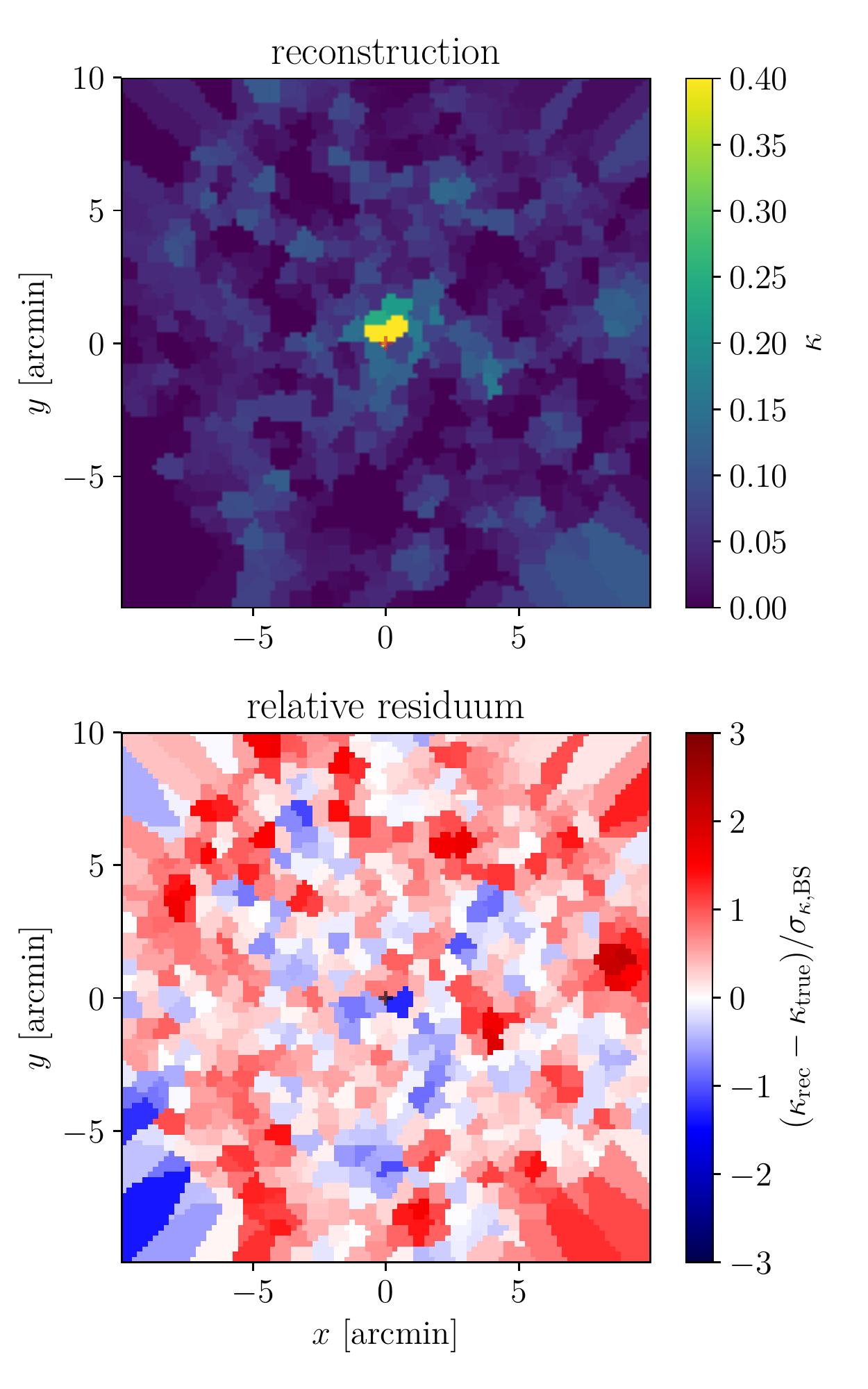}
\caption{weak lensing and X-ray}
\label{fig:nfwmapwxdefbs}
\end{subfigure}
\caption{\textit{top}: Reconstructed convergence map for a NFW mock cluster. \textit{bottom}: relative residuum, compared to error estimate based on bootstrapping.}
\end{figure*}

We ran 500 bootstrap realisations of the mock data, as described in Section \ref{sec:errgof}, to estimate errors on the resulting convergence maps. Figure~\ref{fig:nfwkappaprofdefBSerr} shows the radial convergence profiles of the default reconstructions with error bars based on these realisations. Figures~\ref{fig:nfwmapwdefbs} and \ref{fig:nfwmapwxdefbs} depict the convergence maps and residuals for the shear-only (W from here on) and combined (WX) cases, respectively.

We find the following:
\begin{itemize}
\item The correct amplitude of the central convergence peak is only recovered in the combined reconstruction, but not in the shear-only reconstruction;
\item The bootstrap-based error estimates for both cases are similar;
\item The error estimates increase towards the central peak, where picking a slightly different node position has the largest effect due to a large convergence gradient;
\item The residua relative to the bootstrap error, $\left(\kappa_\mathrm{rec}-\kappa_\mathrm{true}\right)/\sigma_{\kappa,\mathrm{BS}}$, show better agreement in the combined case, where they indicate a fairly unbiased reconstruction (i.e. the mean of the relative residua approximately vanishes).
\end{itemize}

\subsection{Hydrodynamical simulation}
\label{sec:hydrsim}
To investigate the performance of the reconstruction framework when confronted with more realistic cluster morphologies and physics, we turn to a full hydrodynamical simulation of a cluster embedded in the cosmic web. 
We selected a massive cluster from Box2b/hr of the \textit{Magneticum}\footnote{\url{http://www.magneticum.org/}} simulation suite \citep{dolag:inprep.,hirschmann:2014}  through the public web interface\footnote{\url{https://c2papcosmosim.uc.lrz.de}} first presented in \citet{ragagnin:2017}.
The simulations, and our subsequent treatment, adopt a WMAP7 cosmology \citep{komatsu:2011}. Box2b/hr follows $2\times 2880^3$ particles in a volume of $(640 h^{-1}\,\mathrm{Mpc})^3$ with the $N$-body/SPH code \texttt{P-Gadget3} \citep{springel:2005, beck:2016} also incorporating non-gravitational effects such as radiative cooling, heating, star formation, and AGN feedback. 

\begin{figure}
\centering
\includegraphics[width=\hsize]{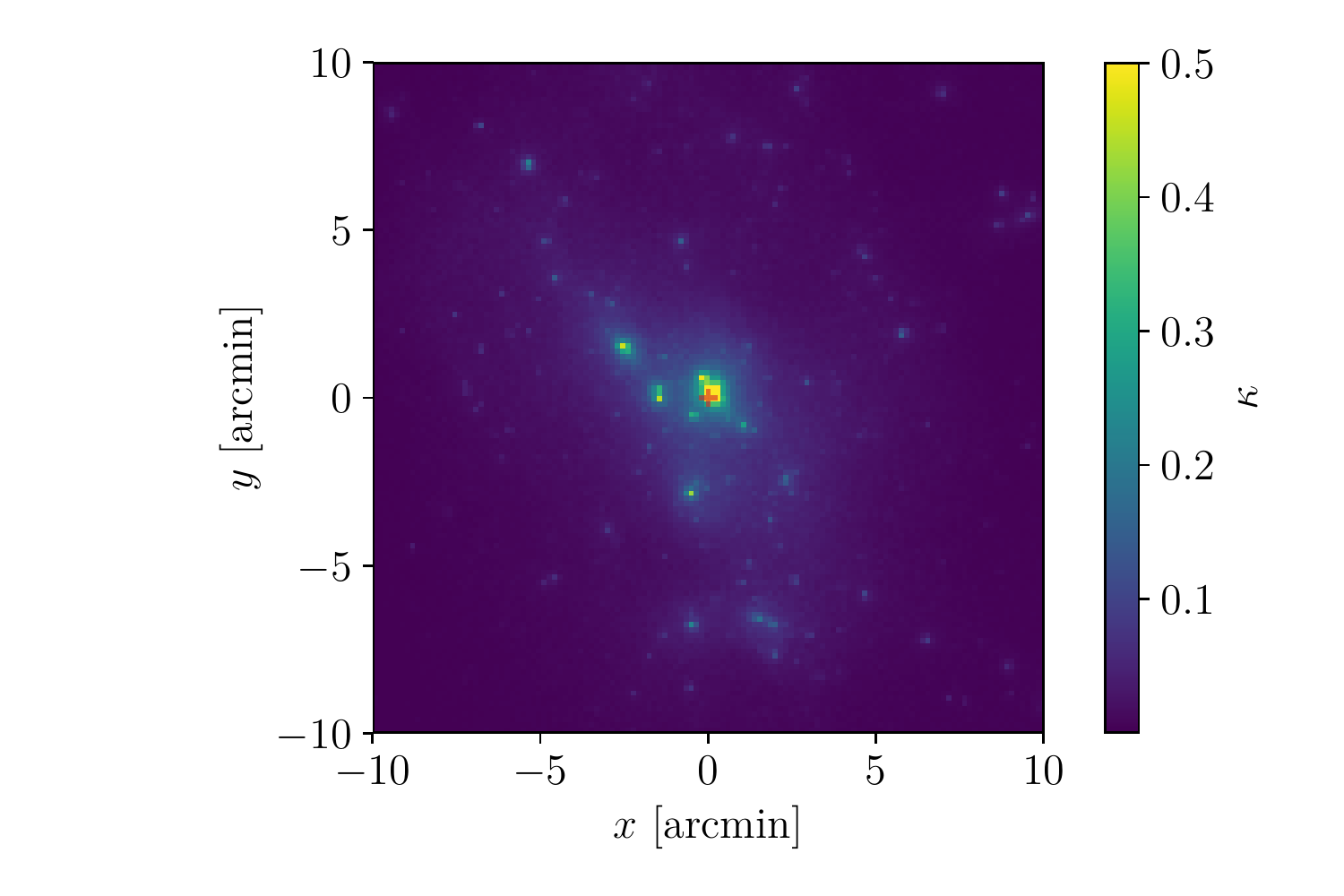}
\caption[Hydrosimulation $\kappa$ map]{Convergence map of the realistic simulated cluster. It clearly features substructure and deviates from spherical symmetry.}
\label{fig:hydrokappatrue}
\end{figure}

We selected a cluster with mass\footnote{$M_{500}$ is the mass enclosed in a spherical region where the mean density is 500 times the background density.} $M_{500} = 5.54\times 10^{14}h^{-1}\, \Msolar$ at redshift $z=0.25$. Its convergence map is shown in Fig.~\ref{fig:hydrokappatrue}. Table~\ref{tab:hydro} lists some of its properties. We selected this particular cluster because it displays a structured and non-spherical morphology while being among the more massive halos in the box, providing sufficient signal-to-noise in the mock lensing and X-ray observations.

\begin{figure}
\centering
\includegraphics[width=\hsize]{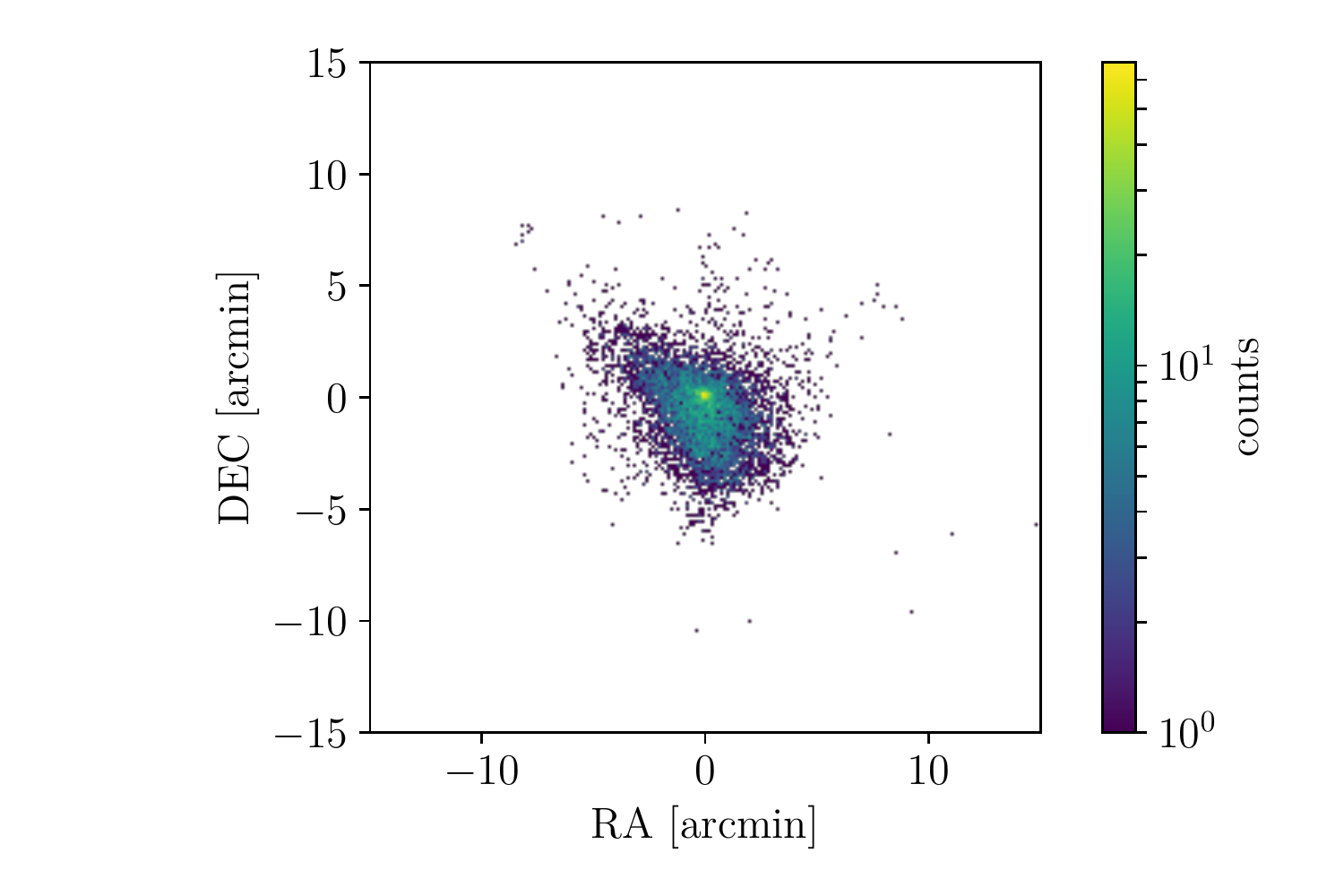}
\caption[XMM count map for realistic cluster]{ Mock XMM count map for a realistic cluster.}
\label{fig:hydroxcount}
\end{figure}

\begin{table}
\centering
\begin{tabular}{c|c}
\hline \hline
Halo & Lensing \\ 
\hline 
$M_{500} = 5.54\cdot10^{14} \Msolar h^{-1}$ & FoV$=(21.3\,\mathrm{arcmin})^2$ \\
$R_{500} =  1.14\,\mathrm{Mpc}$ & $z_\mathrm{s}=1.0$ \\
$z_\mathrm{l}=0.25$& $n_\mathrm{gal}\approx 25\,\mathrm{arcmin}^{-2}$ \\
$T_{500}= 5.98\,\mathrm{keV}$ & $\sigma=0.3$ \\
$f_\mathrm{gas}=0.13$ & ~ \\
\hline 
\end{tabular}
\caption{Defining parameters for the realistic simulated cluster and the mock lensing observations of it.}
\label{tab:hydro}
\end{table}

We employed PHOX \citep{biffi:2012} and SIXTE\footnote{\url{http://www.sternwarte.uni-erlangen.de/research/sixte/}} \citep{schmid:2010}, both available in the  interface, to generate synthetic X-ray observations of $40\,\mathrm{ks}$ exposure with XMM-Newton's EPIC-pn instrument. We used SMAC \citep{dolag:2005a}, to obtain a total surface density map, from which we computed a convergence map and then a shear via the relation
\begin{equation}
\gamma (\bm\theta) = \frac{1}{\pi} \int \mathrm{d}^2\theta' D(\bm \theta - \bm \theta') \kappa(\bm\theta'),
\end{equation}
with the convolution kernel
\begin{equation}
D(\bm \theta) = \frac{-1}{\left(\theta_1 - \mathrm{i} \theta_2\right)^2}.
\end{equation}
We then generated a map of the reduced shear $g$ and sampled it at a source density of $n_\mathrm{gal}\approx 25\,\mathrm{arcmin}^{-2}$ to generate a catalogue of weakly lensed sources at redshift $z_\mathrm{s}=1$, adding again Gaussian shape noise with $\sigma=0.3$.

\begin{figure}
\centering
\includegraphics[width=\hsize]{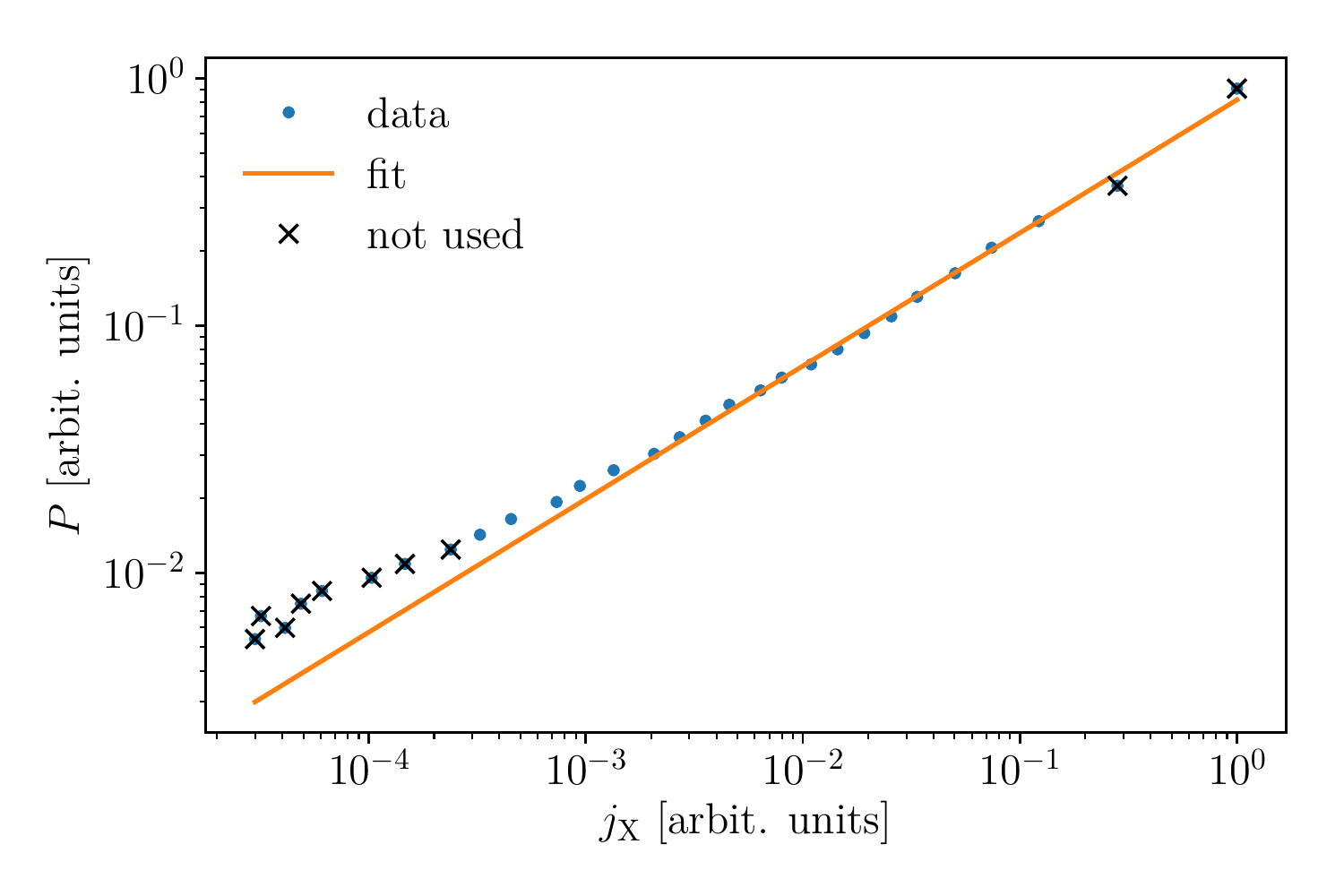}
\caption[Fitting the polytropic index]{Pressure vs.  emissivity in the realistic cluster. The data show a broken power law behaviour indicating that the simple assumptions going into our method are not valid throughout. We restrict the analysis and the fit to obtain the polytropic index to the radial region where the $P$-$j_\mathrm{X}$ follows a straight line in double-logarithmic scaling;}
\label{fig:hydrogammafit}
\end{figure}

We also created a map of the thermal SZ Compton-$y$ signal of the cluster to estimate the polytropic index and a map of the ICM temperature to estimate the sound speed.
For this we plot the electron pressure obtained from deprojecting the Compton-$y$ profile against the X-ray emissivity. If the assumptions discussed in Section \ref{sec:rlxray} hold, both are related by
\begin{equation}
P (r) \propto j_\mathrm{X} ^{2\gamma/(3+\gamma)}(r).
\label{eq:gammafit}
\end{equation}
Figure \ref{fig:hydrogammafit} shows the pressure versus the emissivity together with a fit based on Eq.~\eqref{eq:gammafit}, excluding regions where the simple power law relation seems broken and the assumptions likely do not hold. The best fitting polytropic index is $\gamma = 1.10$. 

As in the previous example, we created $20\,000$ Poisson realisations of the count profile\footnote{The radial bin size is chosen well above the PSF of the X-ray instrument.}, which were deprojected, converted to realisations of the Newtonian potential, and subjected to the same sanity checks. The results of these checks and further diagnostic plots can be found in Appendix \ref{ap:test}. The potential profiles were then projected, employing the analytic continuation scheme up to a radius of four times the truncated data radius. The resulting $\bar\psi_\mathrm{X}$ realisations were used to estimate the corresponding inverse covariance matrix.

\begin{figure}
\centering
\includegraphics[width=\hsize]{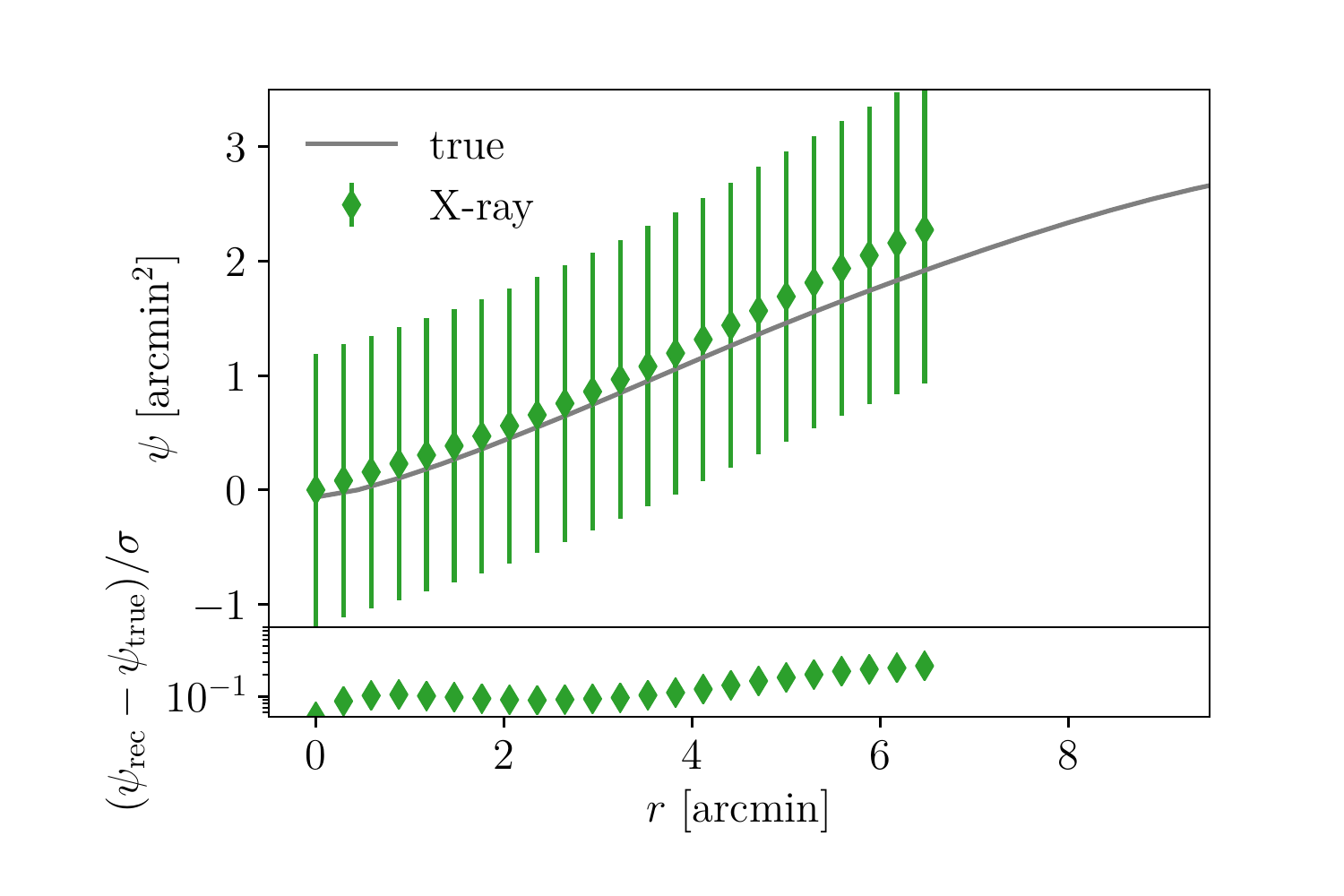}
\caption[Hydrosimulation projected $\psi$]{Lensing potential of a realistic cluster, obtained by projecting the X-ray-based estimate of the Newtonian potential, compared to the true profile.}
\label{fig:hydro_projpsi}
\end{figure}

The mean lensing potential profile obtained from the X-ray data is compared to the actual lensing potential of the simulated cluster in Fig.~\ref{fig:hydro_projpsi}. The curves agree well up to a radius of roughly $5\,$arcmin where the X-ray-based estimate starts to deviate, but still well within the error margins. We chose the cut-off radius based on ballpark estimates of the virial radius from the surface brightness data and experience from Section \ref{sec:art}. The error bars in Fig.~\ref{fig:hydro_projpsi} are noticeably larger than in the NFW case above. One reason for this is the choice of cut-off radius. The further out the extrapolation is taken, the more noisy features, even in the truncated potential profile, influence the continuation and thus the projection and consequently the variance throughout the realisations increases. A balance between extrapolating far enough to counter artefacts while still maintaining constraining power is required. So far rough estimates have proven to work sufficiently well.

\begin{table}
\centering
\begin{tabular}{l|ccc}
\hline \hline
Nodes (nominal) & $300$ & $550$ & $800$\\ 
Nearest neighbours & $40$ &$22$ & $15$\\
Regularisation strength &$200$&$400$&$600$\\
\hline 
\end{tabular} 
\caption[Hydrosimulation reconstruction parameters]{List of parameters defining the outer loop iterations for the \textsc{SaWLens2} reconstruction of the realistic cluster. The regularisation strength is set to be the same for  the regularisation against the convergence and for the regularisation against the shear.}
\label{tab:hydro_rec}
\end{table}

\begin{figure}
\centering
\includegraphics[width=\hsize]{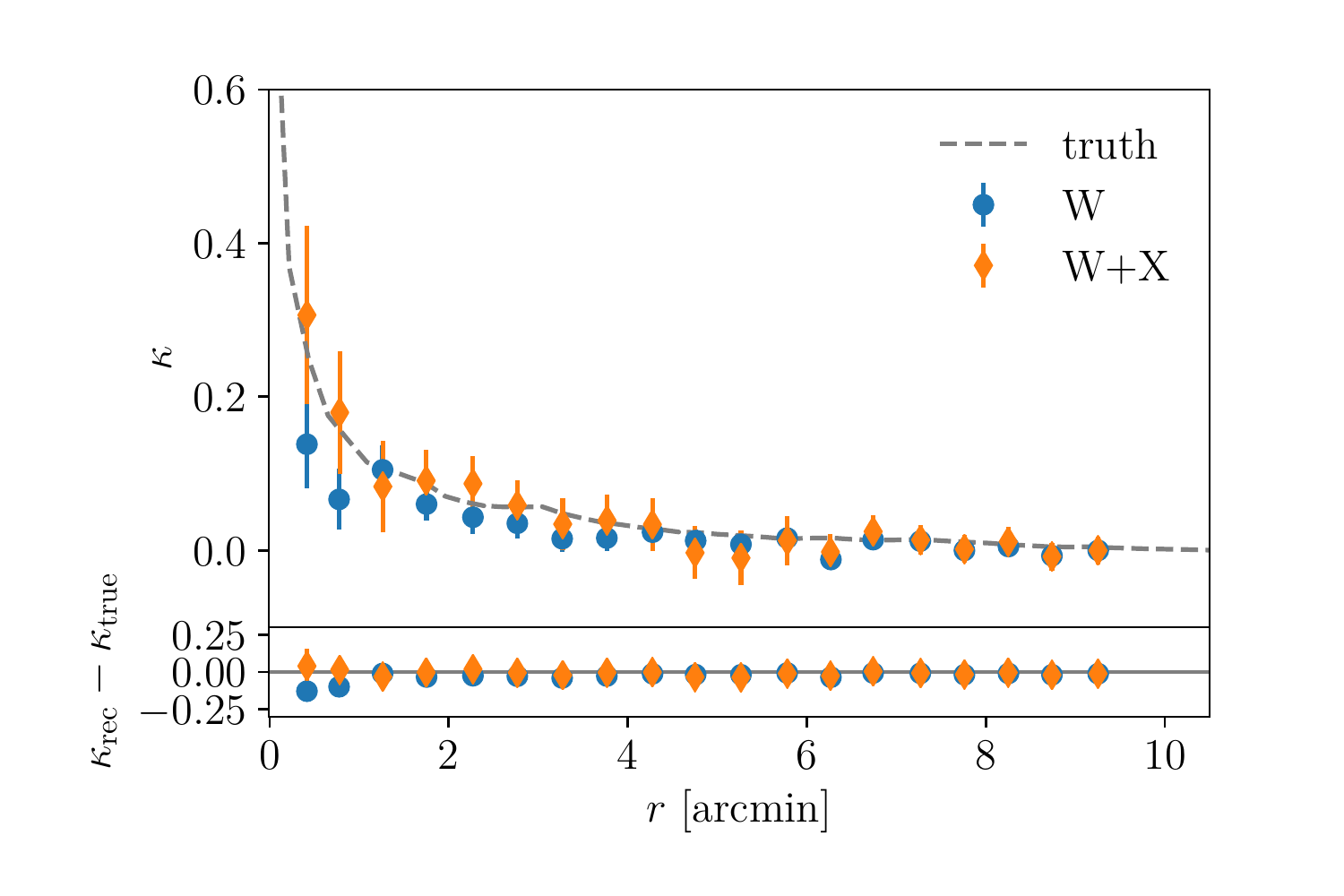}
\caption{Reconstructed convergence profiles for a realistic cluster, with errors inferred from averaging over bootstrapped samples. All curves shifted to coincide in the outermost bin in order to compensate for a slight bias towards higher convergence values.}
\label{fig:hydrokappaprofdefBSerr}
\end{figure}

\begin{figure*}
\begin{subfigure}[b]{0.49\hsize}
\centering
\includegraphics[width=0.9\hsize]{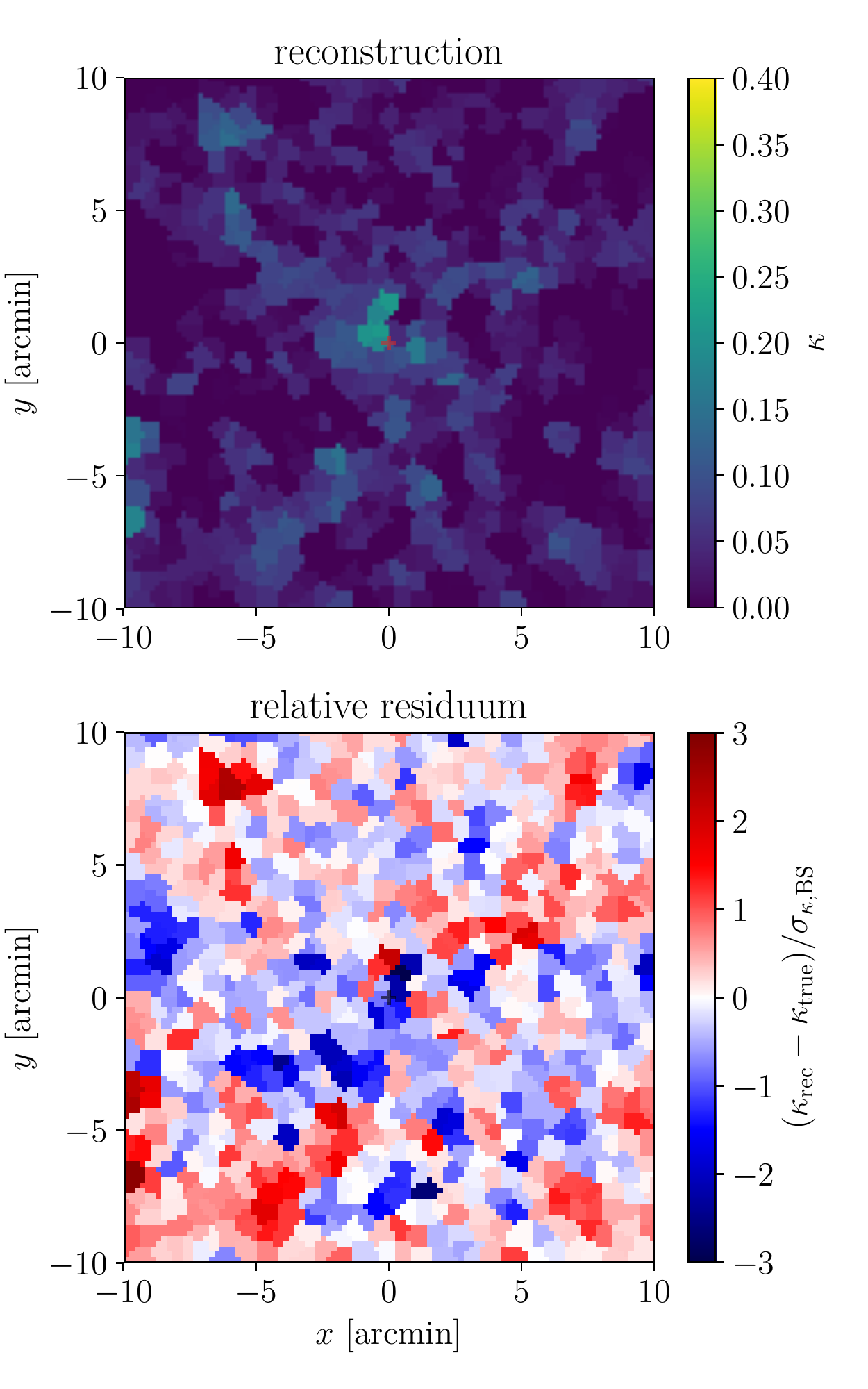}
\caption{weak lensing alone}
\label{fig:hydromapwdefbs}
\end{subfigure}
~
\begin{subfigure}[b]{0.49\hsize}
\centering
\includegraphics[width=0.9\hsize]{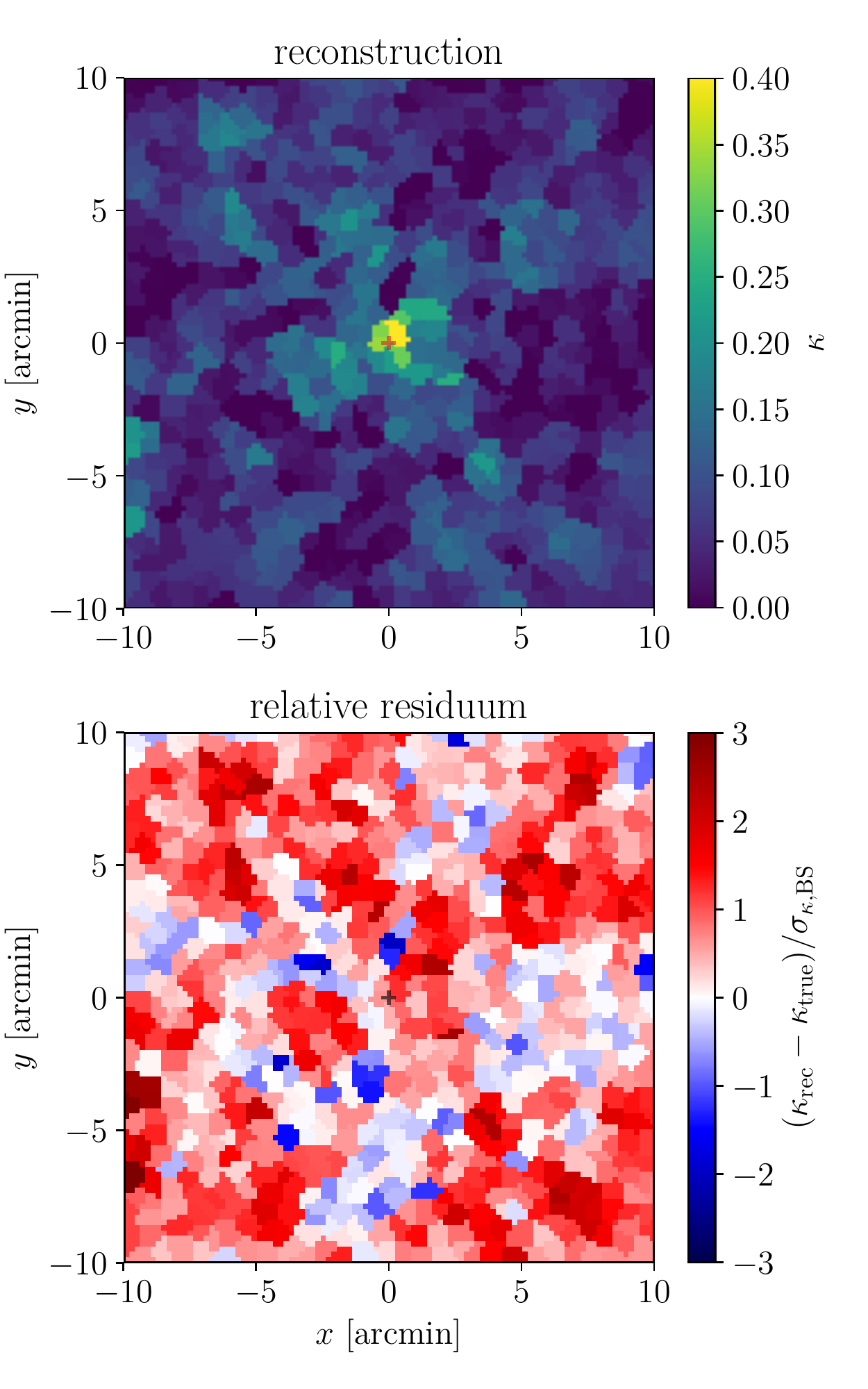}
\caption{weak lensing and X-ray}
\label{fig:hydromapwxdefbs}
\end{subfigure}
\caption{(\textit{top}) Reconstructed convergence map for a realistic cluster. (\textit{bottom}) Relative residuum, compared to error estimate based on bootstrapping.}
\end{figure*}

The following \textsc{SaWLens2} reconstructions, as well as 500 bootstraps, were set up using the parameters collected in Table~\ref{tab:hydro_rec}. Again, we performed one reconstruction  using the shear catalogue alone and one combining shear and X-ray data. Figure~\ref{fig:hydrokappaprofdefBSerr} shows the convergence profiles with errors based on the bootstraps. The resulting convergence maps together with their relative residua based on bootstrapping are shown in Figs.~\ref{fig:hydromapwdefbs} and \ref{fig:hydromapwxdefbs}.

As in the previous use case, the combined reconstruction yields a more pronounced central convergence peak where the radial convergence profile nicely traces the true convergence, especially in the bootstrapped mean. We note, however, that the combined reconstruction is biased high in convergence, as can be seen in the relative residuum map in the lower plot of Fig.~\ref{fig:hydromapwxdefbs}. Experiments with different cut-off radii in the projection step for the X-ray constraint show that the bias is not related to deviations in the projected potential curvature alone. The scaling of the lensing potential is well reproduced (see Fig.~\ref{fig:hydro_projpsi}), and  the combination of shear- and X-ray-based constraints in \textsc{SaWLens2} does not generally introduce a significant bias in the reconstructed convergence (see  previous section). Further investigation reveals that this bias originates in the combination of two factors. The recovered profile of the lensing potential based on X-ray data $\bar\psi_\mathrm{X}$ deviates slightly from the truth (see Fig.~\ref{fig:hydro_projpsi}), and the covariance matrix $\mathcal{C}_\mathrm{X}$ is very dense, effectively correlating every radial bin with every other, as can be seen from the Pearson correlation matrix depicted in Fig.~\ref{fig:hydrocorrmat}. This way, slight overestimations of the radial curvature in the outer bins lead to slight additive biases in the convergence of the joint reconstruction. The radial profiles in Fig.~\ref{fig:hydrokappaprofdefBSerr} are adjusted for this bias. 

In order to quantify how the reconstruction deals with the pronounced triaxial morphology of the cluster, we calculated the second moment tensor
\begin{equation}
E = \sum_n \kappa_n \Theta\left(\kappa_n -\kappa_\mathrm{thresh}\right) \bm\theta_n \otimes \bm\theta_n\label{eq:2ndmom}
\end{equation}
for the convergence map, where the sum runs over all nodes in the reconstruction, and the Heaviside function $\Theta(\kappa_n -\kappa_\mathrm{thresh})$ ensures that only nodes with convergence above a certain threshold are taken into consideration. The eigenvectors of this tensor align with the major axes of the halo, while the ratio of its eigenvalues reflects the square of the ratio of the major axes.   

\begin{figure}
\centering
\includegraphics[width=0.9\hsize]{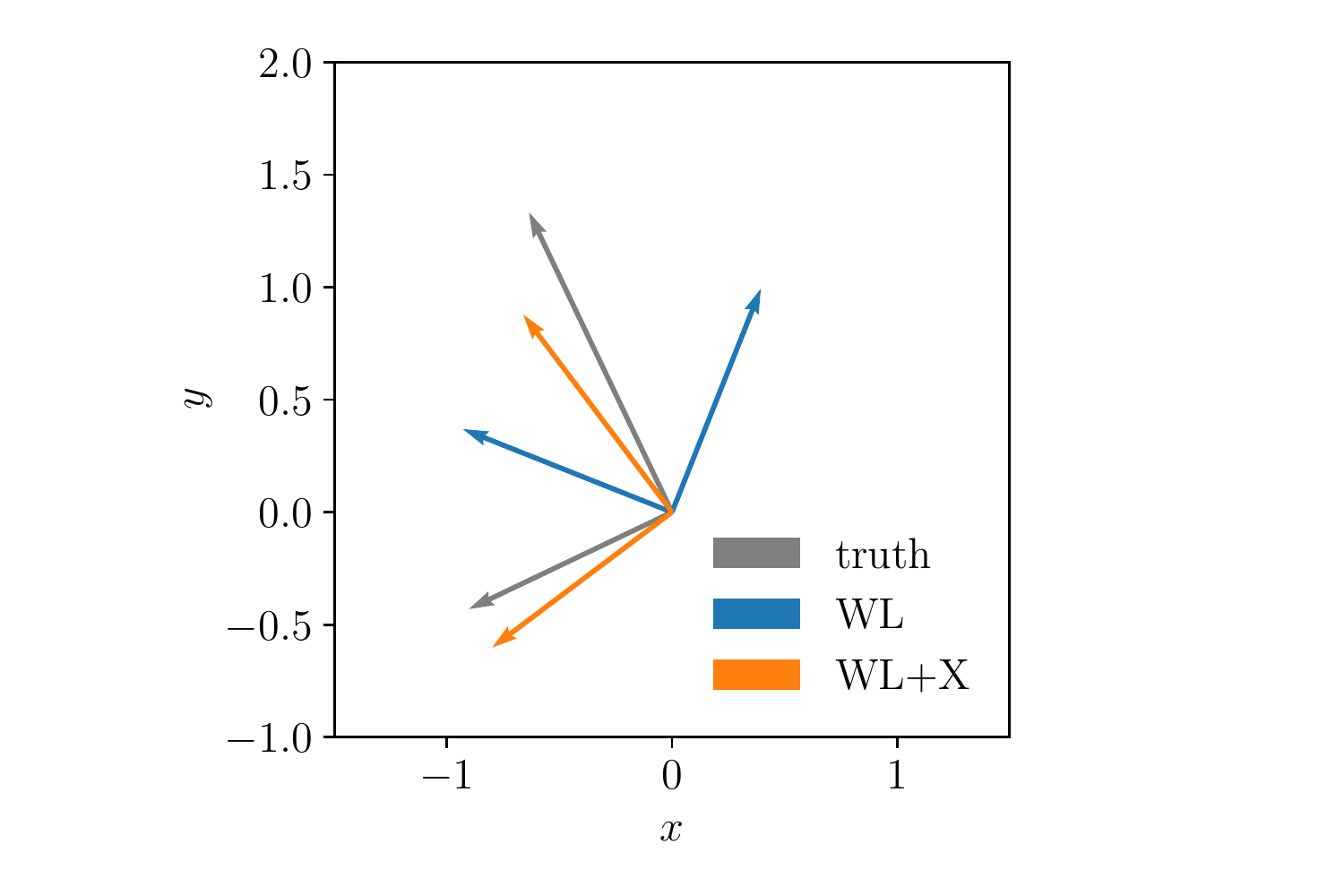}
\caption[Hydrosimulation ellipticity eigenvectors]{Eigenvectors of the second-moment tensor for the reconstructions of a realistic cluster, compared to those based on the true convergence map. The combined case (orange) shows better agreement with true orientations of the cluster than the shear-only case (blue), but both fail to recover the axis ratio.}
\label{fig:hydroevec}
\end{figure}

The corresponding eigenvectors are shown in Fig.~\ref{fig:hydroevec}, scaled to reflect the reconstructed eigenvalue ratio. The X-ray-based information does not add information on the morphology (as it is based on spherical symmetry to create an estimate of the radial profile of the lensing potential), but since the reconstruction is only required to follow this profile on azimuthal average, it also does not impose any morphology. In cases where the noise in the reconstruction is reduced by the addition of X-ray data, the constraints on the morphology may be improved. In the present case, where we used the median reconstructed convergence as a threshold for Eq.~\eqref{eq:2ndmom}, the reconstructions only poorly reflect the axis ratio of the true map, but the combined reconstruction matches the directions of the major axes significantly better than in the shear-only case. 

This is  not a particularly stable result. Changes in the iteration and regularisation settings may alter the inferred second-moment tensor without significantly changing the normalised data residuals or other reconstruction results. It therefore remains doubtful if the addition of symmetrised X-ray constraints can improve the inference of the morphology of a real cluster. 
We   expect, however,  that X-ray constraints using deprojections with spheroidal symmetry are bound to improve this measure;  Fig.~\ref{fig:hydroxcount} shows a clear elongation of the surface brightness, paralleling that of the underlying mass distribution.

\section{Conclusions}
\label{sec:conc}
\subsection{Summary}
In this work we demonstrated how to jointly incorporate weak lensing and X-ray data in a mesh-free free-form reconstruction framework in order to infer the surface mass distribution of a galaxy cluster. We reviewed how the Richardson-Lucy algorithm can be used to deproject the X-ray surface brightness of the ICM and how the resulting emissivity can be connected to the Newtonian potential. We discussed numerical artefacts in the projected potential that may arise when the usable data range of X-ray observations is approximately the same as or narrower than that of  the virial radius, and we present means to compensate these artefacts. We presented the current implementation of the reconstruction framework, \textsc{SaWLens2}, and how X-ray constraints that have been symmetrised during the de- and reprojection steps can be included without imposing any symmetry on the joint reconstruction. Subsequently, we tested the method on two mock clusters, one a simple NFW halo in hydrostatic equilibrium and one a realistic cluster taken from a cosmological hydrodynamical simulation. These tests showed that the addition of X-ray information can significantly improve the reconstruction, especially in the central region, where the peak in convergence can be recovered much better than in the shear-only case. Quantitatively, the reconstruction based on shear alone underestimates the true radial convergence profile in the central region by $2.2\,\sigma_\mathrm{BS}$ or roughly a factor of $2$, whereas the combined reconstruction traces the true convergence almost perfectly down to a projected radius of around $0.5\,$arcmin, with a residuum of only $0.3\,\sigma_\mathrm{BS}$. The combined reconstruction for a realistic cluster displayed a small bias ($<1\,\sigma_{\kappa,\mathrm{BS}}$, see Appendix \ref{ap:test}) towards higher convergence values. This effect is due to the high degree of correlation between radial bins in the X-ray-based potential estimate and slight deviations in the outer bins. Since this is a purely additive bias in the convergence, it is mitigated relatively easily. This highlights once more the need for the careful treatment of artefacts in the projection step as they can alter the results across the whole domain.

\subsection{Scope and possible extensions}
It is worthwhile to also explicitly summarise the limiting and simplifying assumptions that went into our reconstruction method, and how it might be generalised.
We assume that the $\Lambda$CDM framework is an adequate description of the Universe.\\
In the context of gravitational lensing, we limit ourselves to cases that can be sufficiently described in the single thin lens approximation and assume that intrinsic alignment is a negligible source of systematic error on scales relevant for cluster lensing. More involved lensing scenarios, where intrinsic alignment is taken into account can be implemented by altering the computation of the shear covariance matrix.
In the context of using X-ray data as an additional constraint, we assume that the ICM
is in hydrostatic equilibrium. This is needed to connect the gas structure to the total Newtonian potential. However, corrections for hydrostatic bias \citep[e.g.][]{shi:2016} can  be incorporated. We further assume the ICM to be describable as an ideal gas that is polytropically stratified over a significant radial range.
The systematic effects on the reconstructed potential due to these assumptions are currently investigated and will be described in an upcoming paper (Tchernin et al. in prep).
Specifically, the combination of lensing and non-lensing constraints as presented here only works if general relativity is the assumed theory of gravity as otherwise lensing and the ICM density probe different combinations of the Bardeen potentials, which are
in general not identical  \citep{bardeen:1980}.
Parametrised deviations from general relativity \citep[e.g.][]{zhao:2009} can in principle be incorporated to either allow a consistent reconstruction assuming modified gravity or to test for such deviations.\\
Since the approach presented here for X-ray observations, can also be applied to measurements of the thermal SZ signal and member galaxy kinematics, and since strong lensing features are readily included, a joint analysis has the potential to constrain the mass distribution across all scales accessible to observations.

There are several sources of arbitrariness in the method as presented here, namely the values of hyper-parameters like the regularisation strength and the specifics of the iteration and shear averaging scheme, as well as the choice of cut-off radius in the analytic continuation. While there are reasonable experience-based guidelines for choosing these parameters, there is still considerable leeway and the result may be affected by that, without significantly altering the measures of goodness of fit. 

As our test on simulated data show, the presented way of combining weak lensing and X-ray data relies on relatively high spatial resolution for the two data sets. Future applications to real data are therefore likely pointed observations of individual massive clusters rather than statistical samples from survey data. For the former, the tools and methods presented here may allow a free-form assessment of cluster morphology across a wide range of scales, thus potentially informing statistically more precise methods, for example in the modelling of clusters as strong lenses. However, applications to real data are also likely  complicated by several factors. For instance noisy measurements of ICM temperature as well as correlated and/or non-Poissonian noise in the X-ray surface brightness measurements may affect subsequent estimates of the gravitational potential. 

\begin{acknowledgements}
KH thanks Kerstin Paech and Steffen Hagstotz for many helpful discussions.
KH and SH acknowledge the support of the DFG cluster of excellence ‘Origin and Structure of the Universe’
(\href{http://www.universe-cluster.de/}{www.universe-cluster.de}). JM acknowledges funding from the European Union's Horizon 2020 research and innovation programme under the Marie Sk\l{}odowska-Curie grant agreement No. 664931. CT acknowledges the support of the Deutsche Forschungsgemeinschaft under 
BA 1369 / 28-1 and of the Swiss National Science Foundation under P2GEP2 
159139. Parts of this work were performed using the libastro code library developed by the cosmology group at ITA Heidelberg.
\end{acknowledgements}

\bibliographystyle{aa}
\bibliography{MeineBibZ5.bib}

\begin{appendix}
\section{Further diagnostics}
\label{ap:test}
\begin{figure}[h!]
\centering
\includegraphics[width=\hsize]{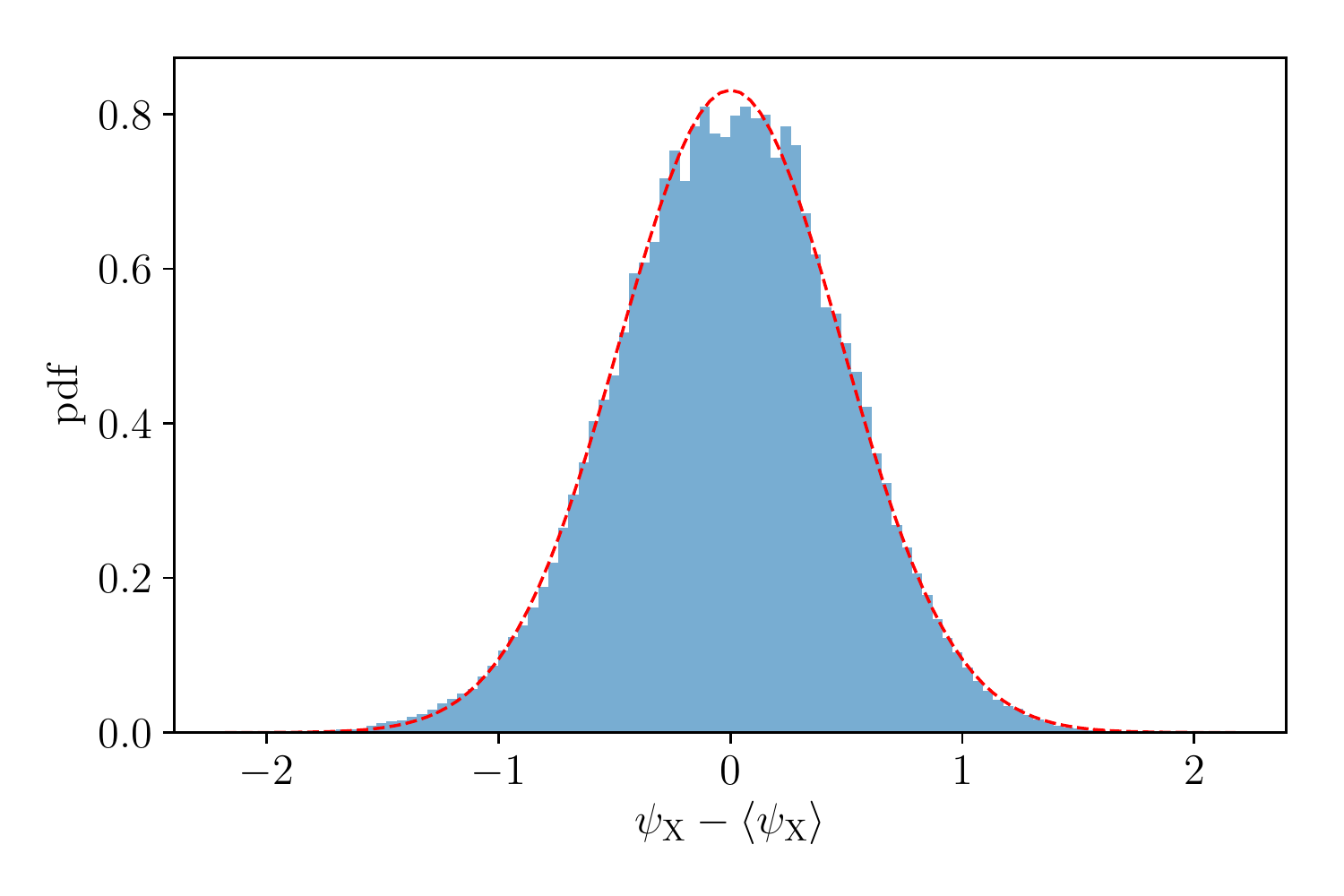}
\caption[NFW $\psi_\mathrm{X}$ realisations are Gaussian distributed]{Histogram of the difference between realisations and the original of the lensing potential obtained from projecting the X-ray-based estimate of the Newtonian potential of a NFW mock cluster. As the overlayed (red) Gaussian shows, the realisations clearly follow a normal distribution.}
\label{fig:nfwpsiXgauss}
\end{figure}

\begin{figure}[h!]
\centering
\includegraphics[width=\hsize]{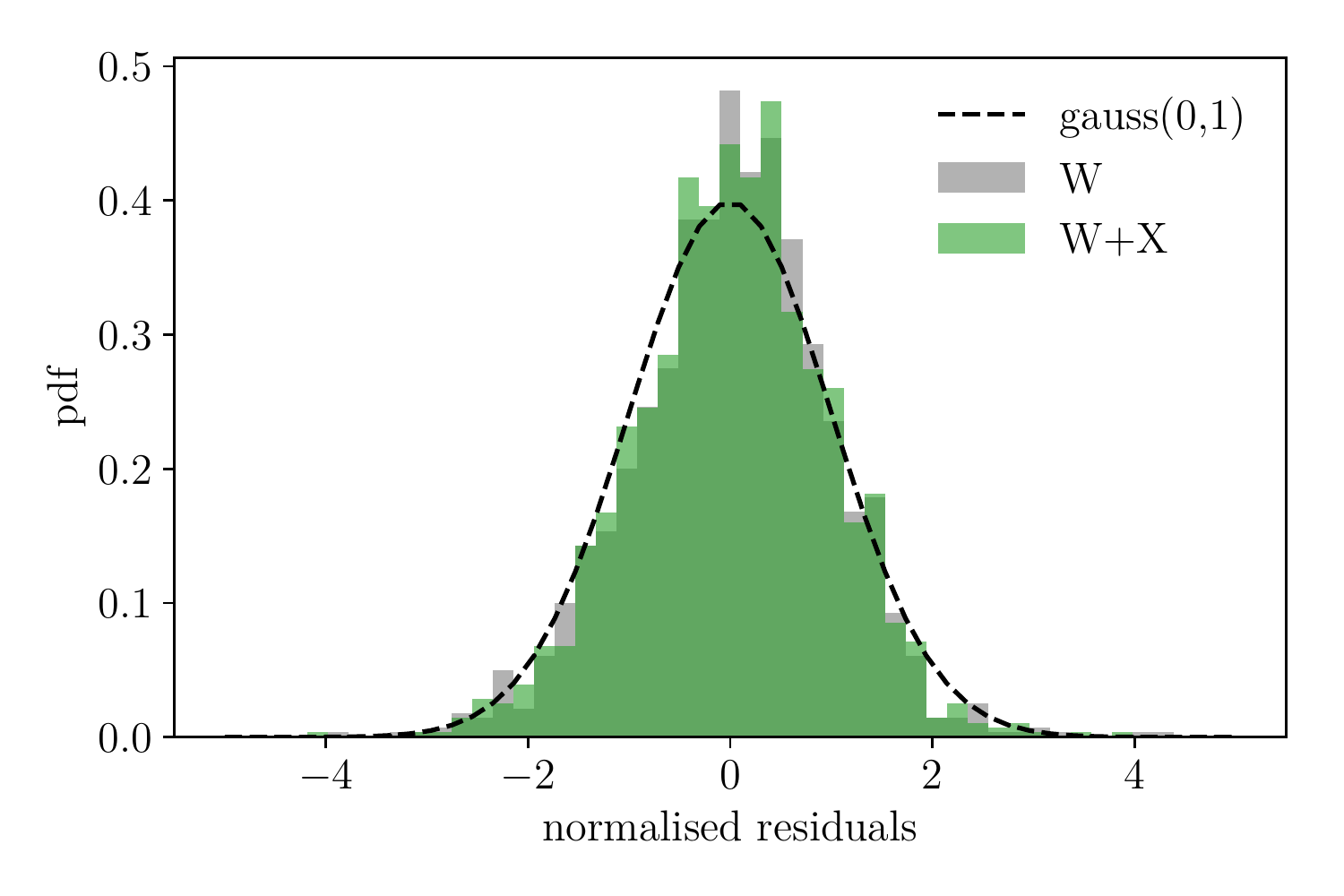}
\caption[normalised residuals, NFW mock]{Histogram of the normalised data residuals in the eigenbasis of the inverse data covariance in the case of a NFW halo. The results of the reconstruction using only shear (grey) and using shear and X-ray data (green) are reasonably consistent with a Gaussian of zero mean and unit variance (dashed line). }
\label{fig:nfwNR}
\end{figure}

\begin{figure}[h!]
\centering
\includegraphics[width=\hsize]{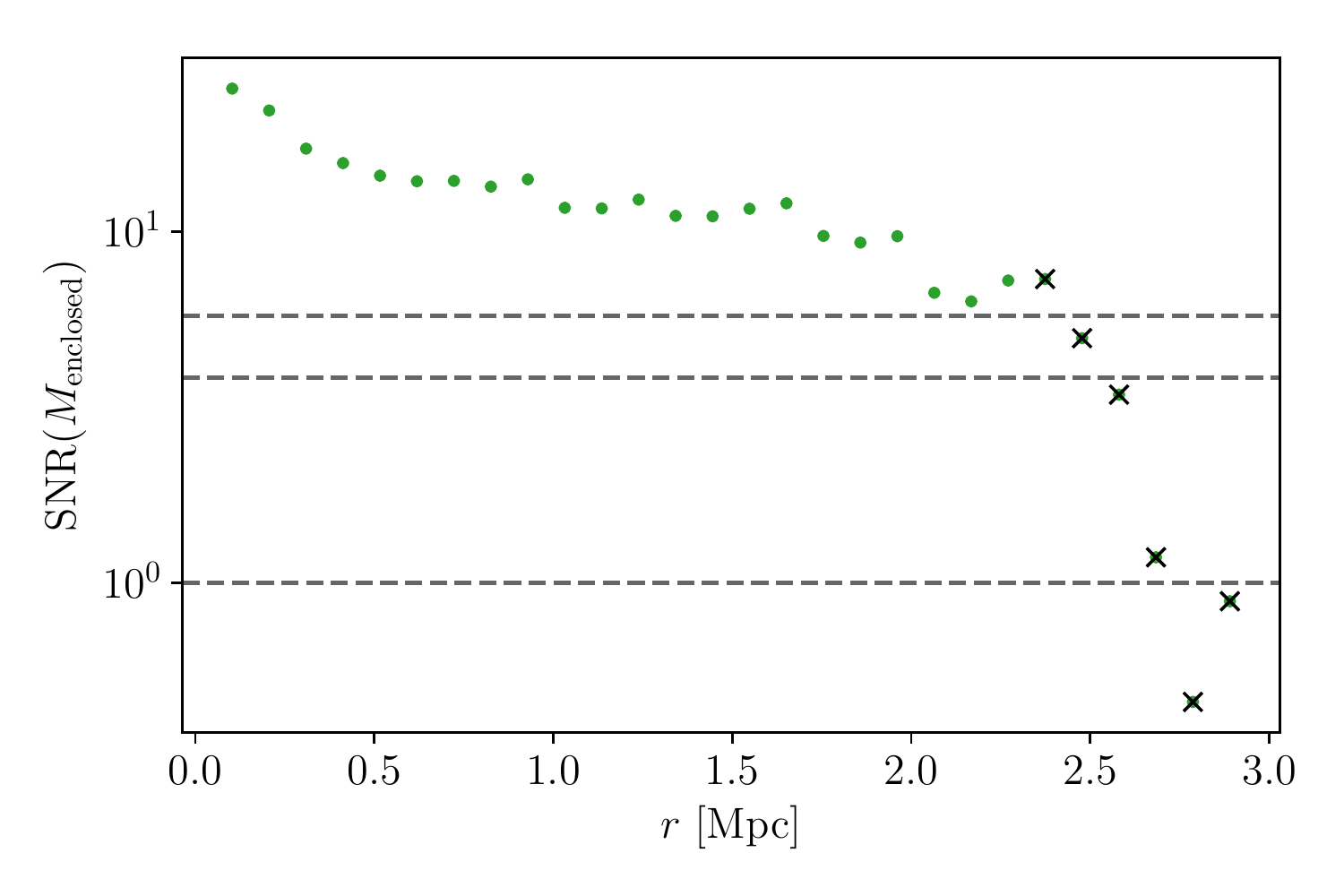}
\caption[Hydrosimulation SNR of enclosed mass]{Signal-to-noise ratio for the enclosed mass as a function of radial bin for a realistic cluster. Horizontal lines denote signal-to-noise values of  1, 3, and 5 respectively. Crossed-out bins fail the sanity checks and are excluded from the analysis.}
\label{fig:hydromsnr}
\end{figure}

\begin{figure}[h!]
\centering
\includegraphics[width=\hsize]{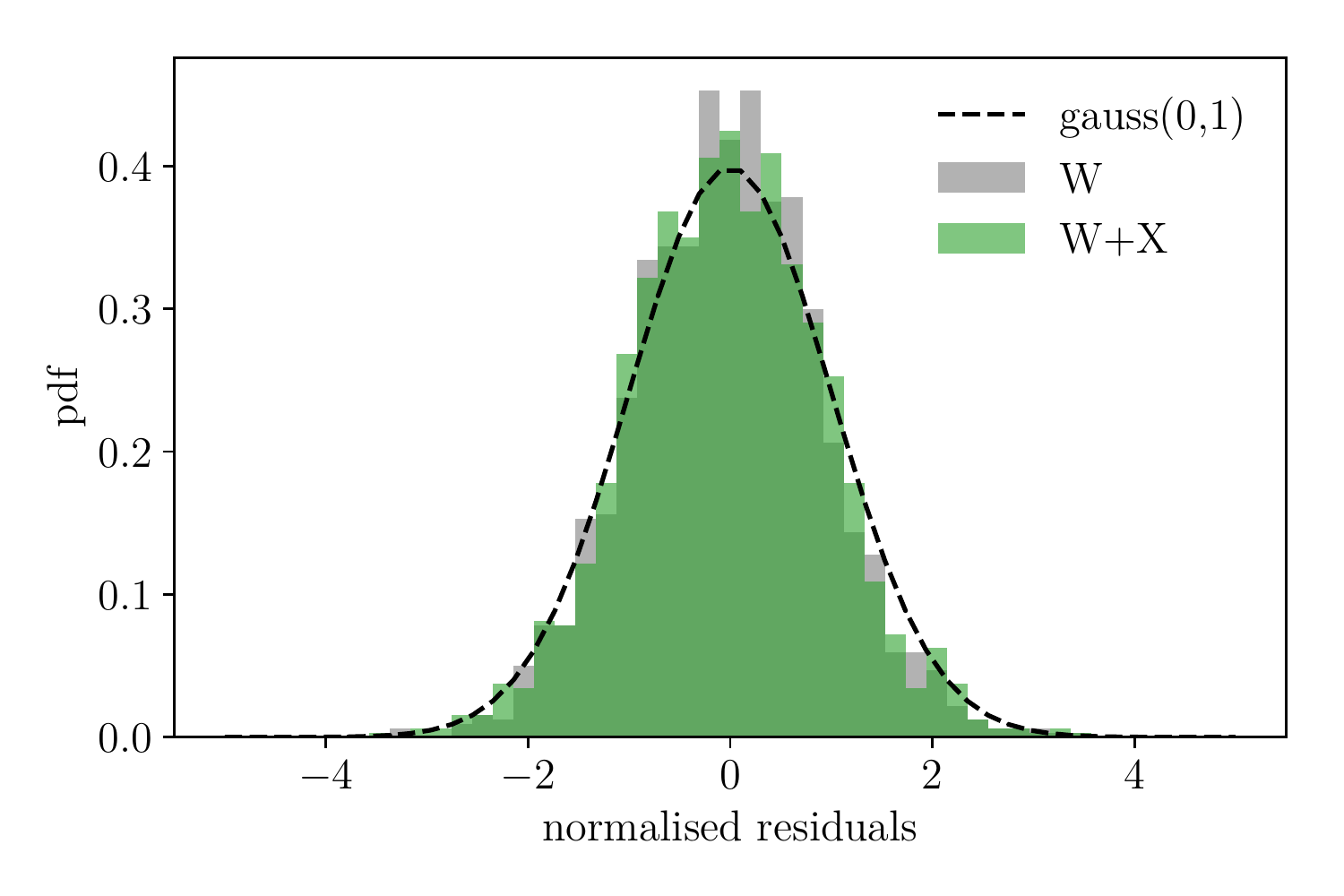}
\caption[normalised residuals, Hydrosimulation]{Histogram of the normalised residuals in the eigenbasis of the data covariance, for the case of a realistic cluster. The results of the reconstruction using only shear (grey)
and using shear and X-ray data (green) are reasonably consistent with a Gaussian of zero mean and unit variance (dashed line).}
\label{fig:hydroNR}
\end{figure}

Figure \ref{fig:nfwpsiXgauss} shows that the assumption of a Gaussian likelihood in Eq.~\eqref{eq:RLchisq} is justified when the noise in the X-ray data is Poissonian. We show results of the sanity checks described in Section \ref{sec:sanch} for the NFW halo in Fig.~\ref{fig:mmsnr} and for the realistic cluster in Fig.~\ref{fig:hydromsnr}, indicating how the recovered profile of the three-dimensional Newtonian potential is truncated before the analytical continuation is applied in the projection step.
Histograms of the normalised residuals of the reconstruction are compared to the expected Gaussian of unit variance in Fig.~\ref{fig:nfwNR} for the NFW halo and Fig.~\ref{fig:hydroNR} for the realistic cluster and show relatively good agreement, indicating appropriately chosen regularisation. 

\begin{figure}[h!]
\centering
\includegraphics[width=\hsize]{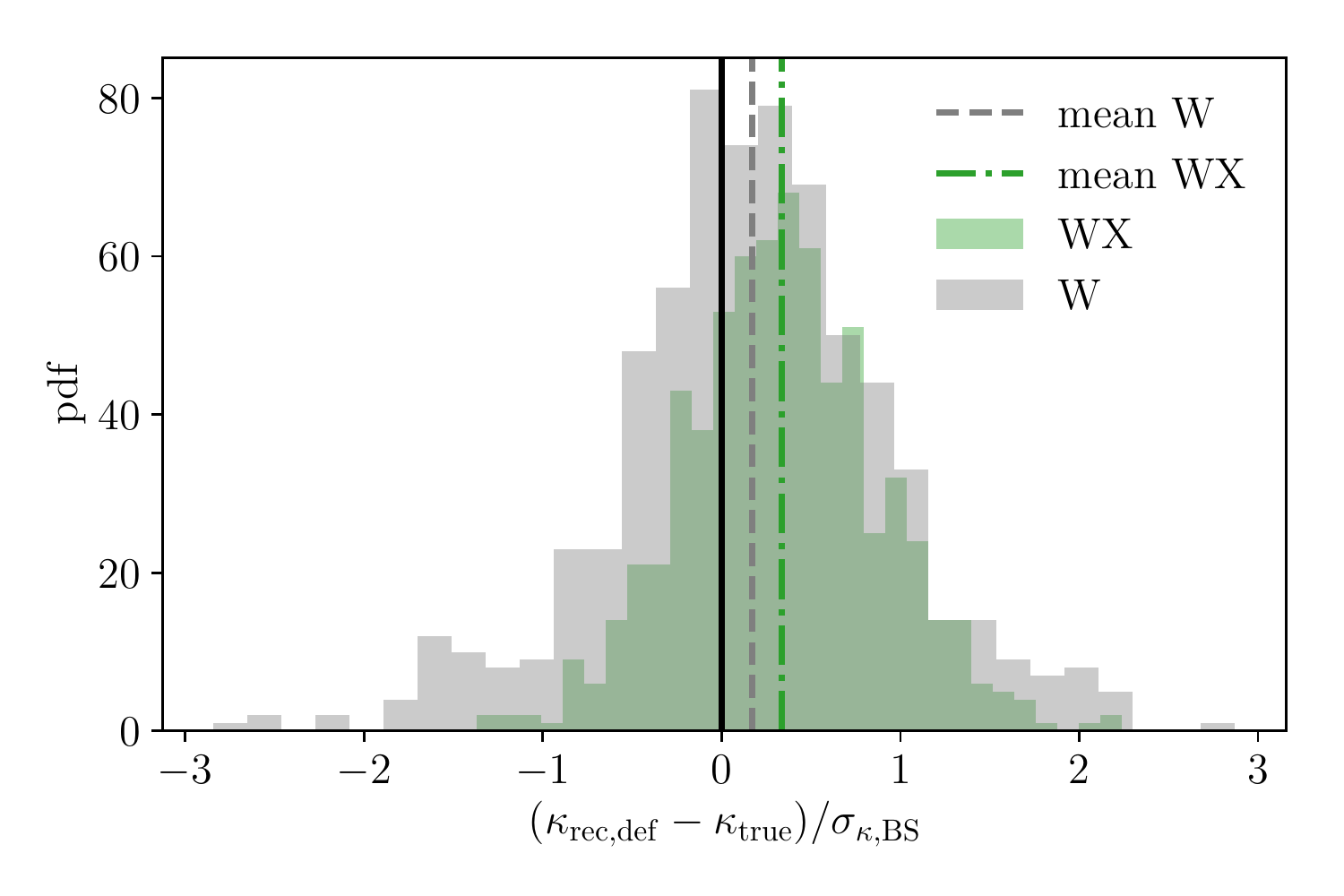}
\caption{Histogram of the deviation of the default reconstructed convergence map of a NFW halo from the true map, in units of the bootstrap-based standard deviation.}
\label{fig:nfw_defrelres}
\end{figure}

\begin{figure}[h!]
\centering
\includegraphics[width=\hsize]{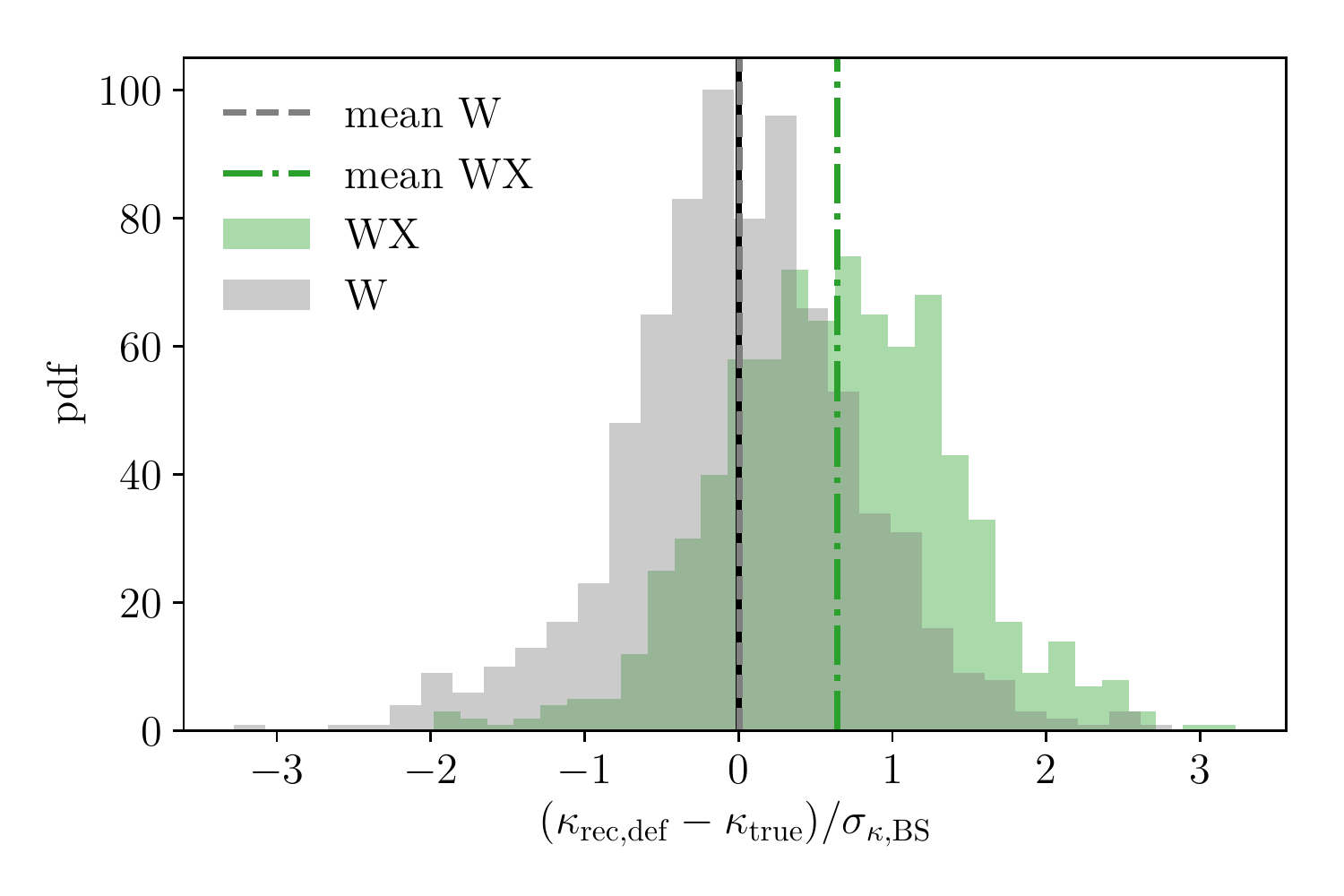}
\caption{Histogram of the deviation of the default reconstructed convergence map of a realistic cluster  from the true map, in units of the bootstrap-based standard deviation.}
\label{fig:hydro_defrelres}
\end{figure}

In Figs.~\ref{fig:nfw_defrelres} and \ref{fig:hydro_defrelres}, the histogram of the relative residuum $\left(\kappa_\mathrm{rec}-\kappa_\mathrm{true}\right)/\sigma_{\kappa,\mathrm{BS}}$ of the reconstruction for the NFW mock and the realistic cluster respectively are shown. While the lensing-only reconstruction does not show any significant bias, the combined reconstruction of the realistic cluster is noticeably shifted towards higher convergence values by about $0.64\,\sigma_{\kappa,\mathrm{BS}}$.

\begin{figure}[h!]
\centering
\includegraphics[width=\hsize]{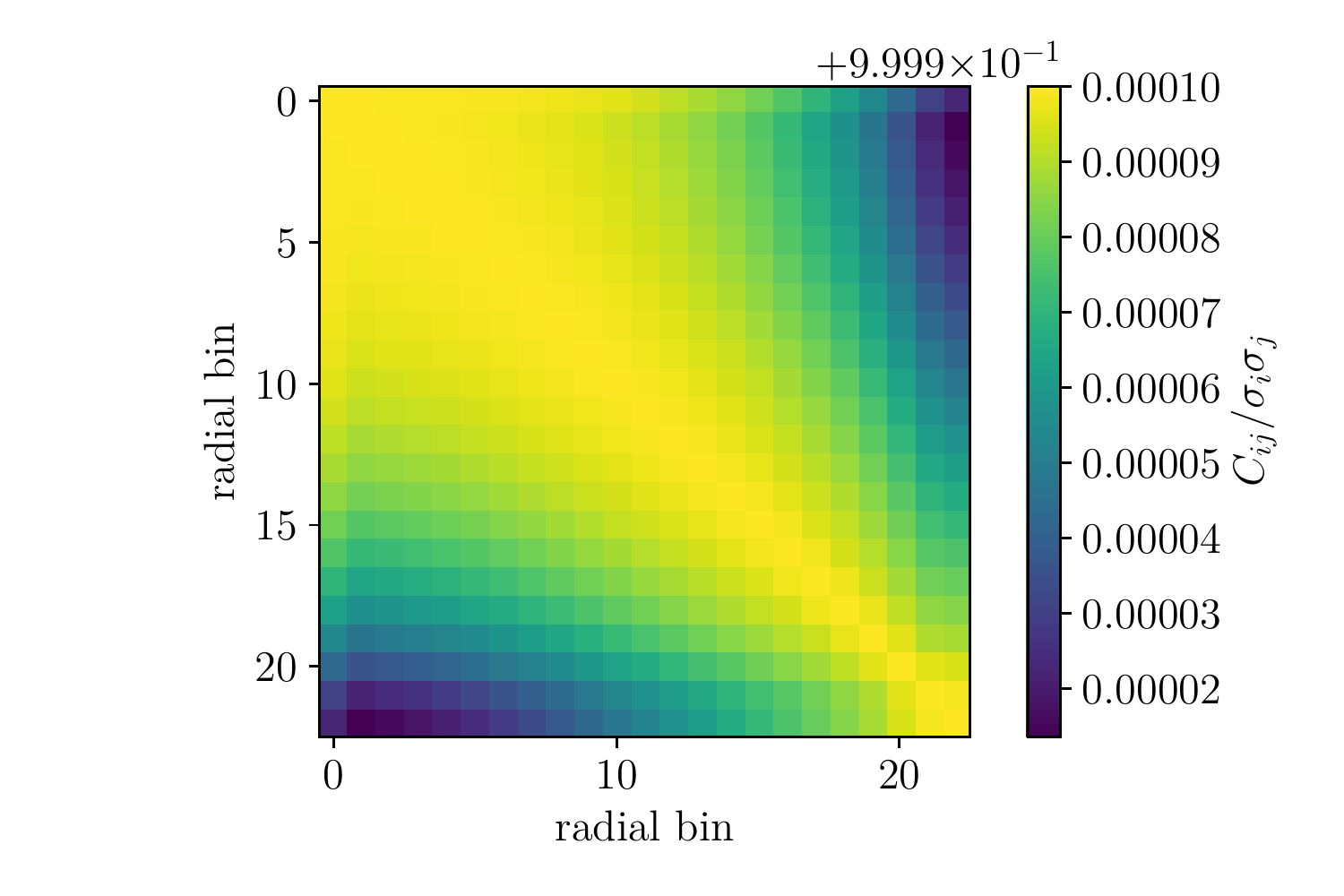}
\caption{The Pearson correlation matrix of the lensing potential profile realisations recovered from X-ray observations of the realistic cluster. It is clear that, due to the projection step, every bin almost completely correlates with every other.}
\label{fig:hydrocorrmat}
\end{figure}

\section{Minimising the X-ray log-likelihood}
For the sake of completeness we explicitly show the contribution of the X-ray data to the LSE to be solved in \textsc{SaWLens2}. All other terms (lensing and regularisation) can be found in \citep{merten:2016} and references therein.

We minimise Eq.~\eqref{eq:RLchisq} with respect to the lensing potential in every node
\begin{equation}
\frac{\partial \chi^2_\mathrm{X}}{\partial \psi_i} \stackrel{!}{=} 0
,\end{equation}
which gives
\begin{eqnarray}
2 \left(O_\mathrm{X}^\mathrm{T} C_\mathrm{X}^{-1} O_\mathrm{X}\right)_{ij} \psi_j &=& 2 \left( O_\mathrm{X}^\mathrm{T} C_\mathrm{X}^{-1} \bar\psi_\mathrm{X} \right)_i\\
B_{ij}^\mathrm{X} \psi_j &=& V_i^\mathrm{X}.
\end{eqnarray}
\end{appendix}

\end{document}